\documentclass{aa}
\usepackage[utf8]{inputenc}
\usepackage[T1]{fontenc}
\usepackage[varg]{txfonts}
\usepackage[english]{babel}
\usepackage{amsmath}
\usepackage{amsfonts}
\usepackage{amssymb}
\usepackage{mathtools}
\usepackage{dcolumn}
\usepackage{booktabs}
\usepackage{multirow}
\usepackage{natbib}
\bibpunct{(}{)}{;}{a}{}{,}
\usepackage{graphicx}
\usepackage{float}
\usepackage{microtype}
\usepackage{hyperref}

\begin{document}
\title{Twisted debris: how differential secular perturbations shape debris disks}
\titlerunning{How differential secular perturbations shape debris disks}
\author{J. A. Sende\inst{1}\and T. Löhne\inst{2}}
\authorrunning{Sende \& Löhne}
\institute{
	Astrophysikalisches Institut, Friedrich-Schiller-Universität Jena, Schillergäßchen~2--3, 07745 Jena, Germany\\
	\inst{1}\email{jan.sende@uni-jena.de}\\
	\inst{2}\email{torsten.loehne@uni-jena.de}
}
\date{Received {\em 4 February 2019}; accepted {\em 13 September 2019}}
\abstract
	{Resolved images suggest that asymmetric structures are a common feature of cold debris disks. While planets close to these disks are rarely detected, their hidden presence and gravitational perturbations provide plausible explanations for some of these features.}%Context
	{To put constraints on the properties of yet undetected planetary companions, we aim to predict what features such a planet imprints in debris disks undergoing continuous collisional evolution.}%Aims
	{We discuss the basic equations, analytic approximations and timescales governing collisions, radiation pressure and secular perturbations. In addition, we combine our numerical model of the collisional evolution of the size and spatial distributions in debris disks with the gravitational perturbation by a single planet.}%Methods
	{We find that the distributions of orbital elements in the disks are strongly dependent on grain sizes. Secular precession is differential with respect to involved semi-major axes and grain sizes. This leads to observable differences between the big grains tracing the parent belt and the small grains in the trailing halo. Observations at different wavelengths can be used to constrain the properties of a possible planet.}%Results
	{}%Conclusions
\keywords{circumstellar matter -- planetary systems -- planet-disk interactions -- celestial mechanics -- methods: analytical -- methods: numerical}
\maketitle

\section{Introduction}\label{sec:introduction}
Despite its relatively low total mass, dust in circumstellar debris disks is readily detected because of its high specific cross section. Current far-infrared detection rates are as high as $\sim$ 20\,\% for nearby stars \citep{eiroa+2013,montesinos+2016,sibthorpe+2018}, indicating how ubiquitous these disks are. Akin to the Edgeworth-Kuiper belt in the Solar system, debris disks are often found at distances of tens or more than one hundred au from their host stars \citep[e.g.][]{rhee+2007,pawellek+2014,matra+2018}. At these distances, they are typically the only detectable components of planetary systems. While thousands of planets have already been detected close to the stars, detections near the cold outer disks are scarce and limited to young, high-mass planets and sub-stellar companions \citep{maldonado+2012}. The old Jupiters and Neptunes that dominate the outer Solar System remain hidden around other stars. The most promising way to characterise these is indirect: through their gravitational perturbations of debris disks.

Gravitational perturbation can lead to substantial orbit changes, even far from the perturber. Mean-motion resonances, typically combined with drifting dust or migrating planets, can produce observable clumps \citep{kuchner+holman2003,wyatt2003,ozernoy+2000}. Secular perturbation can modify orbital eccentricities and orientations \citep{murray+dermott2000}. The latter is a potential explanation for eccentric, off-set disks such as those imaged around Fomalhaut \citep{kalas+2005,stapelfeldt+2004,boley+2012,macgregor+2017}, HR 4796A \citep{schneider+1999,moerchen+2011,kennedy+2018a}, and HD 202628 \citep{schneider+2016}.

The pure gravitational dynamics are well understood. Analytic approximations in the Laplace--Lagrange formalism describe the energy-conserving orbital precession caused by mild perturbations \citep[see, e.g.,][]{murray+dermott2000}. When an orbit is perturbed by a single planet or companion, eccentricity and longitude of pericentre, as well as inclination and longitude of the ascending node oscillate at one frequency. This frequency is determined by the perturber's distance and mass. The perturber modulates shape and orientation of the orbit, without changing the orbital period.

\begin{figure}
	\resizebox{\hsize}{!}{\includegraphics{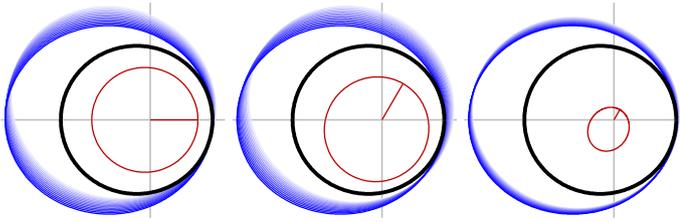}}
	\caption{Three different interior perturbers in \textit{red} that can produce the same eccentric belt (\textit{thick black}) from an initially circular belt. Orbits that deviate from the belt will be sheared by differential precession, as exemplified by orbits of older halo particles (\textit{lighter blue}) trailing that of freshly produced ones (\textit{darker blue}). The location of the star is marked by grey cross hairs. The perturber's periapsis is indicated by \textit{red} lines.\label{fig:idea}}
\end{figure}

Observed belt eccentricities and widths can be related to orbit parameters and masses of unseen perturbers \citep[e.g.][]{rodigas+2014,thilliez+maddison2016}, but this approach provides no unique solutions. There is always a degeneracy between perturber orbit orientations, eccentricities, and distances. Due to typical evolution timescales of several $\mathrm{Myr}$ it is impossible to observe more than a snapshot of a given system. Further degeneracy comes from the fact that the observed narrowness of some belts can have different reasons. On the one hand, a belt could already be equilibrated, corresponding to orbits that are randomly distributed about the precession cycle. In that case, the precession amplitude, that is, the free elements, must be small. On the other hand, precession could have started recently, being still far from equilibrium. These two cases are discussed in \citet{loehne+2017}. The non-equilibrium case is what we focus on throughout the following sections. Figure~\ref{fig:idea} illustrates the resulting degeneracy for a system with a narrow belt.

What is more, the observed dust is always affected by radiation pressure. Small dust grains, which dominate the surface brightness at short wavelengths, are placed on wider orbits by stellar radiation pressure and populate a halo that surrounds the belt of bigger parent bodies \citep{strubbe+chiang2006,krivov+2006,thebault+wu2008}. Depending on where in an eccentric belt these small grains are released, they may stay bound or become unbound, stay closer to the parent belt or populate wider orbits. The asymmetry in the parent belt translates to an asymmetry in the halo \citep{lee+chiang2016}.

The collisional cascade changes the picture again. In our recent work on these eccentric disks \citep{loehne+2017}, we could show how collisions smoothen or remove observable structure. The parent belt appears broadened at short wavelengths, the halo more symmetric. \citet{kim+2018} show that the amplitude of the azimuthal flux asymmetry in eccentric disks depends on the shape of the dust size distribution, which in turn is affected by details of the underlying collision physics. This dependence on material strength and dynamical excitation adds further degeneracy.

In this paper we show and discuss how differential precession shapes belts and halos. Precession timescales depend on the distance between perturber and perturbee, and even on the particles' susceptibility to radiation pressure. The timescales can be very different across a disk, resulting in an effective sheer. This sheer can help reduce the degeneracy and constrain the perturber's orbit. The blue halos in Fig.~\ref{fig:idea} illustrate this idea.

Section~\ref{sec:differential_secular_perturbations} repeats the key points of secular perturbation theory, and amends the formulas to include the dependence on particle size. Furthermore, differential precession is explained, and the relevant timescales are discussed. In Sect.~\ref{sec:implementation} the implementation of perturbations into our existing code \textit{ACE} (Analysis of Collisional Evolution) is described, and the occurring numerical dispersion is discussed. Moreover, the results of a numerical simulation are shown and explained using differential precession. In Sect.~\ref{sec:discussion} the relevance and implications of our results are discussed. The findings are summarised in Sect.~\ref{sec:summary}.

\section{Differential secular perturbations}\label{sec:differential_secular_perturbations}
Secular perturbation theory describes the orbital parameter evolution of bodies under mutual gravitational influence, orbiting the same central object. In the following an abridged version of secular perturbation theory is presented. Corrections are made to include the influence of radiation pressure, which introduces a size dependence for smaller particles. A model describing a differentiation in the orbital evolution of particles created by collisions in the debris disk is derived and put into context.

\subsection{Secular perturbation theory}\label{sec:secular_perturbation_theory}
While eccentricity $e$, longitude of pericentre $\varpi$, inclination $i$ and longitude of the ascending node $\Omega$ can change, the semi-major axis $a$ is not modified by secular perturbations. In the following, we will focus on the evolution of $e$ and $\varpi$,\footnote{Inclination and longitude of the ascending node are currently not included in our simulation due to computation time constraints.} giving a short recapitulation of \citet{murray+dermott2000}.

In order to simplify the equations, the eccentricities $e$ and the longitudes of pericentres $\varpi$ of all bodies are transformed into the $h$--$k$ coordinate system:
\begin{align}
	\label{eq:hkdefinition}
	h=e\sin{\varpi}\text{,} && k=e\cos{\varpi}\text{.}
\end{align}
This transformation simplifies the maths involved and helps interpreting the results. In the $h$--$k$ space, the complicated changes of $e$ and $\varpi$ become operations on the vectors $(h,k)$. 

We are interested in the case of one planet and a low-mass debris disk orbiting around a star. If the planet is much more massive than the debris disk, only the orbital elements of the debris are significantly changed. The orbital elements of the planet are assumed constant because the gravitational influence of the disk on the planet's orbit is small. However, in a multi-planet system the mutual gravitational influence of the planets would lead to a change in orbital elements over time.

In the case of a single planet, only the differential equations of the debris need to be solved:
\begin{align}
	\label{eq:hkdiff}
	\dot{h}= A k+A_\mathrm{p} k_\mathrm{p}\text{,} && \dot{k}=-A h-A_\mathrm{p} h_\mathrm{p}\text{.}
\end{align}
The constants $A$ and $A_\mathrm{p}$ are calculated using the mean motion $n_\mathrm{K}=\sqrt{G M_\star/a^3}$,\footnote{In the following we mark the mean motion with $n_\mathrm{K}$ and the phase space density with $n$.} the mass ratio $\epsilon=m_\mathrm{p}/M_\star$, the semi-major axis functions $\alpha$ and $\bar{\alpha}$, and the Laplace coefficients $b_{3/2}^{(1)}\!\left(\alpha\right)$ and $b_{3/2}^{(2)}\!\left(\alpha\right)$:
\begin{align}
	\label{eq:Acoefficients}
	A           = \frac{n_\mathrm{K}}{4}\epsilon\,\alpha\,\bar{\alpha}\,b_{3/2}^{(1)}\!\left(\alpha\right)\text{,} && A_\mathrm{p} =-\frac{n_\mathrm{K}}{4}\epsilon\,\alpha\,\bar{\alpha}\,b_{3/2}^{(2)}\!\left(\alpha\right)\text{.}
\end{align}
The semi-major axis functions depend on the position of the debris relative to the planet: If the semi-major axis of the planet $a_\mathrm{p}$ is smaller than the axis of the debris $a$ we have the case of an \textit{inner perturber} ($a_\mathrm{p}<a$). If the semi-major axis of the planet $a_\mathrm{p}$ is larger than the axis of the debris $a$, we have the case of an \textit{outer perturber} ($a_\mathrm{p}>a$). If the semi-major axes of the planet and the debris are similar ($a_\mathrm{p}\approx a$), secular perturbation theory fails to provide a good approximation:
\begin{align}
	\label{eq:alphas}
	\alpha = 
	\begin{cases} 
	 a_\mathrm{p}/a           & \text{if } a_\mathrm{p} < a\quad \text{(inner perturber)}\\
	 a          /a_\mathrm{p} & \text{if } a_\mathrm{p} > a\quad \text{(outer perturber)}
	\end{cases}\text{,}\nonumber\\
	\bar{\alpha} = 
	\begin{cases} 
	 1                       & \text{if } a_\mathrm{p} < a\quad \text{(inner perturber)}\\
	 a          /a_\mathrm{p} & \text{if } a_\mathrm{p} > a\quad \text{(outer perturber)}
	\end{cases}\text{.}
\end{align}

Inserting $A$ and $A_\mathrm{p}$ from equation~(\ref{eq:Acoefficients}) into equation~(\ref{eq:hkdiff}), the general solution for the orbit evolution is:
\begin{alignat}{2}
	\label{eq:hkforced}
	h&=e_\mathrm{forced}\sin{\varpi_\mathrm{p}}&+e_\mathrm{free}\sin{\left(\frac{2\pi}{T_\mathrm{prec}^\mathrm{full}}t+\phi_\mathrm{free}\right)}\text{,}\nonumber\\
	k&=e_\mathrm{forced}\cos{\varpi_\mathrm{p}}&+e_\mathrm{free}\cos{\left(\frac{2\pi}{T_\mathrm{prec}^\mathrm{full}}t+\phi_\mathrm{free}\right)}\text{.}
\end{alignat}
The solution can be interpreted as the orbit of the debris precessing with a constant angular velocity $A_\mathrm{prec}=2\pi/T_\mathrm{prec}^\mathrm{full}$ around the forced eccentricity vector $(h_\mathrm{forced},k_\mathrm{forced})=(e_\mathrm{forced}\sin{\varpi_\mathrm{p}},e_\mathrm{forced}\cos{\varpi_\mathrm{p}})$ in a constant distance $e_\mathrm{free}$. 

The forced eccentricity $e_\mathrm{forced}$ and the precession timescale $T_\mathrm{prec}^\mathrm{full}$ depend on the planet's eccentricity $e_\mathrm{p}$, and the semi-major axis functions $\alpha$ and $\bar{\alpha}$:
\begin{align}
	\label{eq:eforced}
	e_\mathrm{forced}=\frac{b_{3/2}^{(2)}\!\left(\alpha\right)}{b_{3/2}^{(1)}\!\left(\alpha\right)}e_\mathrm{p}\text{,} && T_\mathrm{prec}^\mathrm{full}=\frac{2\pi}{A_\mathrm{prec}}=\frac{8\pi}{\alpha\,\bar{\alpha}\,n_\mathrm{K}b_{3/2}^{(1)}\!\left(\alpha\right)}\frac{M_\star}{m_\mathrm{p}}\text{.}
\end{align}
The free eccentricity $e_\mathrm{free}$ and the free angle $\phi_\mathrm{free}$ are determined from the initial conditions. They can be calculated using the starting position in the $h$--$k$ space, and inserting $t=0$ into equation~(\ref{eq:hkforced}), solving for them. 

If the perturber is far away from the disk ($\alpha\ll 1$), the Laplace coefficients can be series expanded and reduced to their leading-order terms: $b_{3/2}^{(1)}\!\left(\alpha\right)\approx 3\alpha$, and $b_{3/2}^{(2)}\!\left(\alpha\right)\approx\frac{15}{4}\alpha^2$. \citep{murray+dermott2000} The forced eccentricity and the precession timescale thus are
\begin{align}
	\label{eq:innerouter}
	e_\mathrm{forced}&\approx\frac{5}{4}\alpha e_\mathrm{p}\text{,}\nonumber\\
	T_\mathrm{prec}^\mathrm{full}&\approx \frac{2\pi}{A_\mathrm{prec}} = \frac{8\pi}{3}\frac{M_\star}{m_\mathrm{p}}n_\mathrm{K}^{-1}\times
	\begin{cases} 
		\alpha^{-2} & \text{(inner perturber)}\\
		\alpha^{-3} & \text{(outer perturber)}
	\end{cases}\text{.}
\end{align}
In a resolved, extended disk, $\alpha$ can be used to distinguish between possible planets because $\alpha$ increases when the semi-major axis of the debris $a$ is closer to the planets axis $a_\mathrm{p}$. Combining equation~(\ref{eq:alphas}) with equation~(\ref{eq:innerouter}), one gets the following proportionalities:
\begin{align}
	\label{eq:proportionalities}
	e_\mathrm{forced}&\propto\begin{cases} 
		a^{-1}\cdot a_\mathrm{p}^{+1}\cdot e_\mathrm{p} \quad \text{(inner perturber)}\\
		a^{+1}\cdot a_\mathrm{p}^{-1}\cdot e_\mathrm{p} \quad \text{(outer perturber)}
	\end{cases}\text{,}\nonumber\\
	T_\mathrm{prec}^\mathrm{full}&\propto\begin{cases} 
		a^{+7/2}\cdot a_\mathrm{p}^{-2}\cdot m_\mathrm{p}^{-1} \quad \text{(inner perturber)}\\
		a^{-3/2}\cdot a_\mathrm{p}^{+3}\cdot m_\mathrm{p}^{-1} \quad \text{(outer perturber)}
	\end{cases}\text{.}
\end{align}
For an \textit{inner perturber}, debris with larger $a$ has a smaller forced eccentricity $e_\mathrm{forced}$ and a longer precession timescale $T_\mathrm{prec}^\mathrm{full}$. The dependency is inverse for an \textit{outer perturber} -- debris with larger $a$ has a larger $e_\mathrm{forced}$ and a shorter $T_\mathrm{prec}^\mathrm{full}$. In both cases, the debris with a semi-major axis closer to the planets axis precesses faster. In the case of an \textit{inner perturber}, the inner disk edge leads in front of the outer edge, whereas in the case of an \textit{outer perturber} it drags behind. Furthermore, $T_\mathrm{prec}^\mathrm{full}$ scales with $\alpha^{-3}$ instead of $\alpha^{-2}$, leading to a slower precession for an outer planet, compared to an inner planet with the same $\alpha$. In the following we focus on the case of an \textit{inner perturber}, but it is possible to derive similar formulas for the \textit{outer one}.

\subsection{Radiation pressure}\label{sec:radiation_pressure}
In the above it is assumed that only gravitational forces act on the debris. In reality however, small dust particles can be heavily influenced by radiation pressure and stellar winds. The above equations are amended, using the \citet{burns+1979} formalism: $F=F_\mathrm{grav}-F_\mathrm{rad}=(1-\beta)F_\mathrm{grav}$, where $\beta$ is the ratio between radiative push and gravitational pull. (For our analytical model, we ignore stellar winds.) Applying the changes to equation~(\ref{eq:innerouter}), it is easy to check, that the forced eccentricity does not change, but the precession timescale now depends on the $\beta$ value of the dust as well:
\begin{align}
	\label{eq:precessiontime}
	e_\mathrm{forced}=\frac{5}{4}\alpha e_\mathrm{p}\text{,} && T_\mathrm{prec}^\mathrm{full}=\frac{8\pi}{3}\frac{M_\star}{m_\mathrm{p}}n_\mathrm{K}^{-1} \sqrt{1-\beta} \alpha^{-2}\text{.}
\end{align}
Dust with the same semi-major axis (and thus same $\alpha$), but higher $\beta$ value precesses faster in the $h$--$k$ phase space than dust with lower $\beta$. Thus, if one knows the position and $\beta$ value of particles in the $h$--$k$ space, one can calculate $e_\mathrm{forced}$ and $T_\mathrm{prec}^\mathrm{full}$.

The $\beta$ value depends on the material and size of the particle, and the luminosity and spectrum of the host star \citep{burns+1979}:
\begin{align}
	\label{eq:beta}
	\beta=\frac{F_\mathrm{rad}}{F_\mathrm{grav}}=\frac{3}{16\pi}\frac{L_\star}{G M_\star}\frac{Q_\mathrm{PR}(s)}{c \rho s}\text{,}
\end{align}
with $L_\star$ and $M_\star$ being the luminosity and the mass of the host star, $\rho$ the dust density, $Q_\mathrm{PR}(s)$ the radiation pressure efficiency, and $s$ the size of the particle. Usually $Q_\mathrm{PR}(s)$ is calculated using Mie theory, and depends on the stellar spectral type and the material used. 

\subsection{Dust release}
The detectable dust of a debris disk is created in mutual collisions of bigger objects, and in cometary activity. These source objects are usually not detectable because of their small specific surface area. Yet, the small dust particles they release, have a much larger specific surface area and are therefore detected.

The orbits of fragments released by collisions or cometary activity depend on several parameters, such as the masses of the colliders, the impact velocity and location, etc. However, if a fragment is released from a larger parent body on a roughly circular orbit, the orbit parameters of the fragment only depend on its $\beta$ value and the semi-major axis of the source $a_\mathrm{src}$ \citep[equation~(9)]{strubbe+chiang2006}:
\begin{align}
	\label{eq:collision}
	\frac{a_\mathrm{src}}{a}\approx2-\frac{1}{1-\beta}=\frac{1-2\beta}{1-\beta}\text{,} && e\approx\frac{\beta}{1-\beta}\text{.}
\end{align}
While these equations are derived for fragment production in a circular parent belt, they still hold for parent belt eccentricities that are small compared to the $\beta$ values of small halo grains, that is, for $e_\mathrm{src} \ll \beta$.

As described in Sect.~\ref{sec:radiation_pressure}, the $\beta$ value depends on the size of the particle, therefore the orbits of collisional fragments depend on the sizes as well. As shown on the left in Fig.~\ref{fig:collision_scheme}, larger fragments, which usually have a $\beta$ close to zero, will stay in nearly circular orbits within the parent belt. Smaller fragments however, due to their larger $\beta$, are on elliptic orbits and create the halo of the disk. Particles close to $\beta\approx1/2$ are on highly eccentric orbits. They constitute the outermost part of the halo.

When launched from circular orbits, particles with $\beta\gtrsim 1/2$ have an eccentricity greater than one, and are therefore on unbound orbits, getting ejected from the system. (Due to their anti-clockwise motion in Fig.~\ref{fig:collision_scheme}, unbound fragments can only be ejected into the \textit{red} shaded region.) Using equation~(\ref{eq:beta}) one finds a minimum particle size for each star, below which particles get unbound. This size is called the blowout size $s_\mathrm{blow}$. For early type stars it is greater than for late type stars:
\begin{align}
	s_\mathrm{blow}=\frac{3}{8\pi}\frac{L_\star}{G M_\star}\frac{Q_\mathrm{PR}(s)}{c \rho}\text{,}
\end{align}
where $Q_\mathrm{PR}$ is the radiation pressure efficiency \citep{burns+1979}. The radiation pressure of low luminosity stars is not strong enough, and thus no small size limit exists \citep{reidemeister+2011}.

\subsection{Differential precession}\label{sec:differential_precession}
Combining the dependence of equation~(\ref{eq:collision}) on $\beta$ with the forced eccentricity and precession timescale of equation~(\ref{eq:precessiontime}), one gets the full $\beta$--dependent phase space evolution of fragments produced in a circular belt:
\begin{align}
	\label{eq:combination}
	e_\mathrm{forced}&=\frac{5}{4}\frac{a_\mathrm{p}}{a_\mathrm{belt}} e_\mathrm{p} \frac{1-2\beta}{1-\beta}\text{,}\nonumber\\
	T_\mathrm{prec}^\mathrm{full}&=\frac{8\pi}{3}\frac{M_\star}{m_\mathrm{p}}n_\mathrm{K}^{-1} \left(\frac{a_\mathrm{belt}}{a_\mathrm{p}}\right)^2 \frac{\left(1-\beta\right)^{4}}{\left(1-2\beta\right)^{7/2}}\text{,}
\end{align}
where $a_\mathrm{belt}$ is the semi-major axis of the initial debris disk belt. According to equation~(\ref{eq:collision}), fragments of different size will be produced at different locations in the $h$--$k$ space. These then precess at different speeds around different forced eccentricities, as described in equation~(\ref{eq:combination}).

\begin{figure*}
	\centering
	\includegraphics[width=17cm]{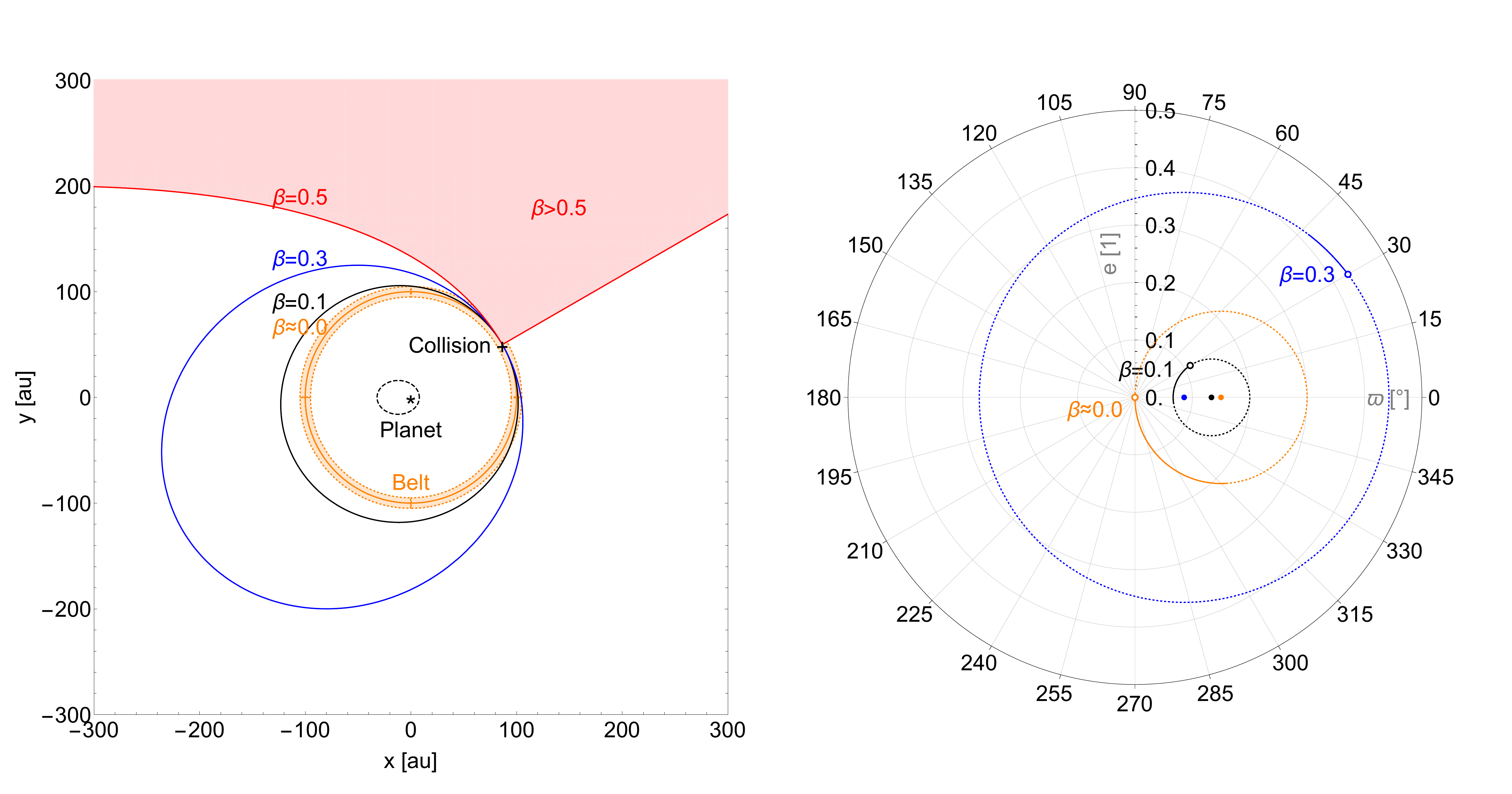}
	\caption{
		\textit{Left}: Initial orbits of fragments with $\beta$ values of $0.0$ (\textit{orange}), $0.1$ (\textit{black}), and $0.3$ (\textit{blue}), emitted in a single collision from a circular orbit at $a=100\mathrm{au}$ (and $\varpi=30^{\circ}$; particles moving anti-clockwise). Particles with $\beta\ge0.5$ (\textit{red}) are on unbound orbits. \textit{Right}: Initial positions (\textit{circle}) of the same fragments in the $h$--$k$ space and their circles of movement (\textit{dashed line}) around the forced eccentricity (\textit{dot} of according colour). The arcs they covered in the $h$--$k$ space within 5~Myr are shown as \textit{solid lines}. Given the planet at $a_\mathrm{p}=20\mathrm{au}$ with $e_\mathrm{p}=0.6$ (and $\varpi_\mathrm{p}=0^{\circ}$), they have precession times of $T^\mathrm{full}_\mathrm{prec}=19.3\mathrm{Myr}$ (for $\beta=0.0$), $T^\mathrm{full}_\mathrm{prec}=27.7\mathrm{Myr}$ (for $\beta=0.1$), and $T^\mathrm{full}_\mathrm{prec}=114.8\mathrm{Myr}$ (for $\beta=0.3$).
	}
	\label{fig:collision_scheme}
\end{figure*}

Fig.~\ref{fig:collision_scheme} illustrates the effect on particles with different $\beta$ values, produced in the same collision. Small particles, which have higher $\beta$ values than large particles, begin their dynamic evolution on more eccentric orbits. They start farther from their forced elements in the $h$--$k$ space and therefore move around wider circles. Small particles have larger semi-major axes as well, leading to lower forced eccentricities and slower precession rates. Although all particles had the same time to precess, large belt particles with $\beta\approx0$ covered a much longer arc in the $h$--$k$ space than the small halo particles.

\begin{figure*}
	\centering
	\includegraphics[width=17cm]{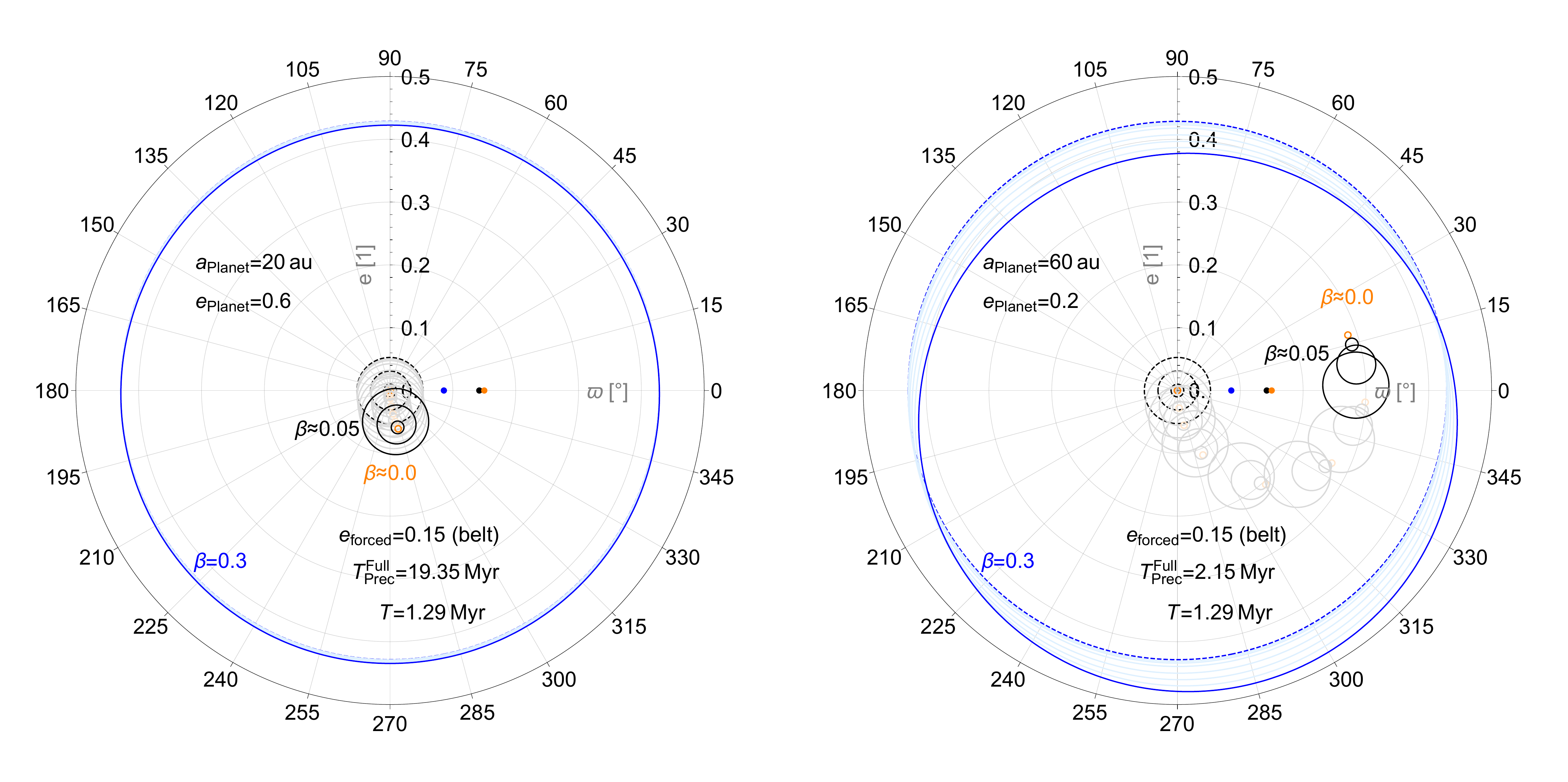}
	\caption{
		$h$--$k$ space evolution of parent belt particles (\textit{orange}; $a_\mathrm{belt}=100\mathrm{au}$; $\beta=0.0$; $e\approx0.0$), and particles created in collisions of parent belt objects; with fragment $\beta$ values of $\beta=0.01;0.03;0.05$ (\textit{black}), and $0.3$ (\textit{blue}). The belt is around a star of $M=1.92M_\odot$, and the collisional fragments are perturbed by a $m_\mathrm{p}=2.5M_\mathrm{Jup}$ mass planet, with $a=20\mathrm{au}$ and $e=0.6$ (\textit{left}); $a=60\mathrm{au}$ and $e=0.2$ (\textit{right}). The initial distribution of particles created in the collisions (\textit{dashed line}), and the position after a time $T$ (\textit{solid line}) are shown. In between steps are marked with faint colours.
	}
	\label{fig:planet_comparison}
\end{figure*}

While Fig.~\ref{fig:collision_scheme} shows the case of particles created in a single collision, Fig.~\ref{fig:planet_comparison} shows how the ensemble of collisional fragments created at different longitudes in the parent belt would evolve. For simplicity, the parent belt is assumed to be an azimuthally uniform, narrow, circular ring. The collisional fragments are initially distributed on circles in the $h$--$k$ space (\textit{black} and \textit{blue}). In realistic belts with finite widths and dynamical excitations, these rings would have a certain width, too.

The parent belt has the smallest semi-major axis, and therefore precesses the fastest in the $h$--$k$ space. Large particles, with low $\beta$ values, stay close to the parent belt because their eccentricity spread is small. The smaller the grains, the higher are the $\beta$ values, and the wider is their eccentricity spread. The small particles are on orbits with large semi-major axes and form the disk halo. Their forced eccentricity is low and they precess at a slower rate compared to the parent belt and the large grains.

Figure~\ref{fig:planet_comparison} additionally shows how the phase space evolution is highly dependent on the planets' semi-major axes. Remembering equation~(\ref{eq:proportionalities}), we see that for both the planet in the \textit{right} panel and the planet in the \textit{left} panel, the forced eccentricity is the same. The precession time of the disks however differs significantly (\textit{left}: $19.35\mathrm{Myr}$; \textit{right}: $2.15\mathrm{Myr}$). After the same evolution time of $T=1.29\mathrm{Myr}$ both disks look very different. However, with the time $T$ usually unknown, the two cases cannot be distinguished based on their secular precession alone. After $T=11.61\mathrm{Myr}$, the distribution on the left would be identical to the one on the right at $T=1.29\mathrm{Myr}$. It is only through the combined action with Poynting-Robertson drag and collisional evolution, acting on different timescales, that this degeneracy is broken.

\subsection{Comparison of timescales}\label{sec:timescales}
In addition to the precession timescale, several other timescales need to be considered for a debris disk: the crossing timescale, the timescale of Poynting--Robertson drag, and the average collision time of particles. In order to make comparisons easier, these timescales are expressed in terms of the physical properties of the host star, the planet, and the debris disk.

The crossing timescale was derived by \citet{mustill+wyatt2009}. They showed that particles which are initially separated by different semi-major axes, are forced on crossing orbits by secular perturbations. This leads to an increased destruction of particles in mutual collisions. The time it takes the debris to get on crossing orbits, is the crossing timescale:
\begin{align}
	T_\mathrm{cross}\approx&1.61\times 10^{-3} \frac{\left(1-e_\mathrm{p}^2\right)^{3/2}}{e_\mathrm{p}}\nonumber\\
	&\left(\frac{a_\mathrm{belt}}{10\mathrm{au}}\right)^{9/2}\left(\frac{M_\star}{M_\odot}\right)^{1/2}\left(\frac{m_\mathrm{p}}{M_\mathrm{Jup}}\right)^{-1}\left(\frac{a_\mathrm{p}}{10\mathrm{au}}\right)^{-3}\mathrm{Myr}\mathrm{.}
\end{align}
Furthermore, we derive a similar form of the Poynting--Robertson timescale for particles created in the initial debris disk. This is the time it takes a dust particle to leave the its initial pericentre position $q=a\left(1-e\right)$ due to Poynting--Robertson drag
\begin{align}
	T_\mathrm{PR}=\frac{q}{\dot{q}}\approx&0.32 \frac{1}{\beta\left(4-5\beta\right)}\frac{\left(1-\beta\right)}{\left(1-2\beta\right)^{3/2}}\nonumber\\
	&\left(\frac{a_\mathrm{belt}}{10\mathrm{au}}\right)^2\left(\frac{M_\star}{M_\odot}\right)^{-1}\mathrm{Myr}\mathrm{.}
\end{align}
In order to allow a better comparison, we rewrite the precession timescale in the same manner:
\begin{align}
	\label{eq:precscale}
	T^\mathrm{full}_\mathrm{prec}\approx&4.42\times 10^{-2} \frac{\left(1-\beta\right)^4}{\left(1-2\beta\right)^{7/2}}\nonumber\\
	&\left(\frac{a_\mathrm{belt}}{10\mathrm{au}}\right)^{7/2}\left(\frac{M_\star}{M_\odot}\right)^{1/2}\left(\frac{m_\mathrm{p}}{M_\mathrm{Jup}}\right)^{-1}\left(\frac{a_\mathrm{p}}{10\mathrm{au}}\right)^{-2}\mathrm{Myr}\mathrm{.}
\end{align}

Lastly, we need the collisional timescales. This is the time it takes a particle on average to be destroyed in another mutual collision. In contrast to the other timescales, the collisional timescale directly depends on the mass of the debris disk. We did not derive an analytic expression, but use the collision timescale calculated in our simulation presented in Sect.~\ref{sec:implementation} instead. (For possible analytical solutions see \citet{wyatt2005} \& \citet{loehne+2008}.)

\begin{figure*}
	\centering
	\includegraphics[width=17cm]{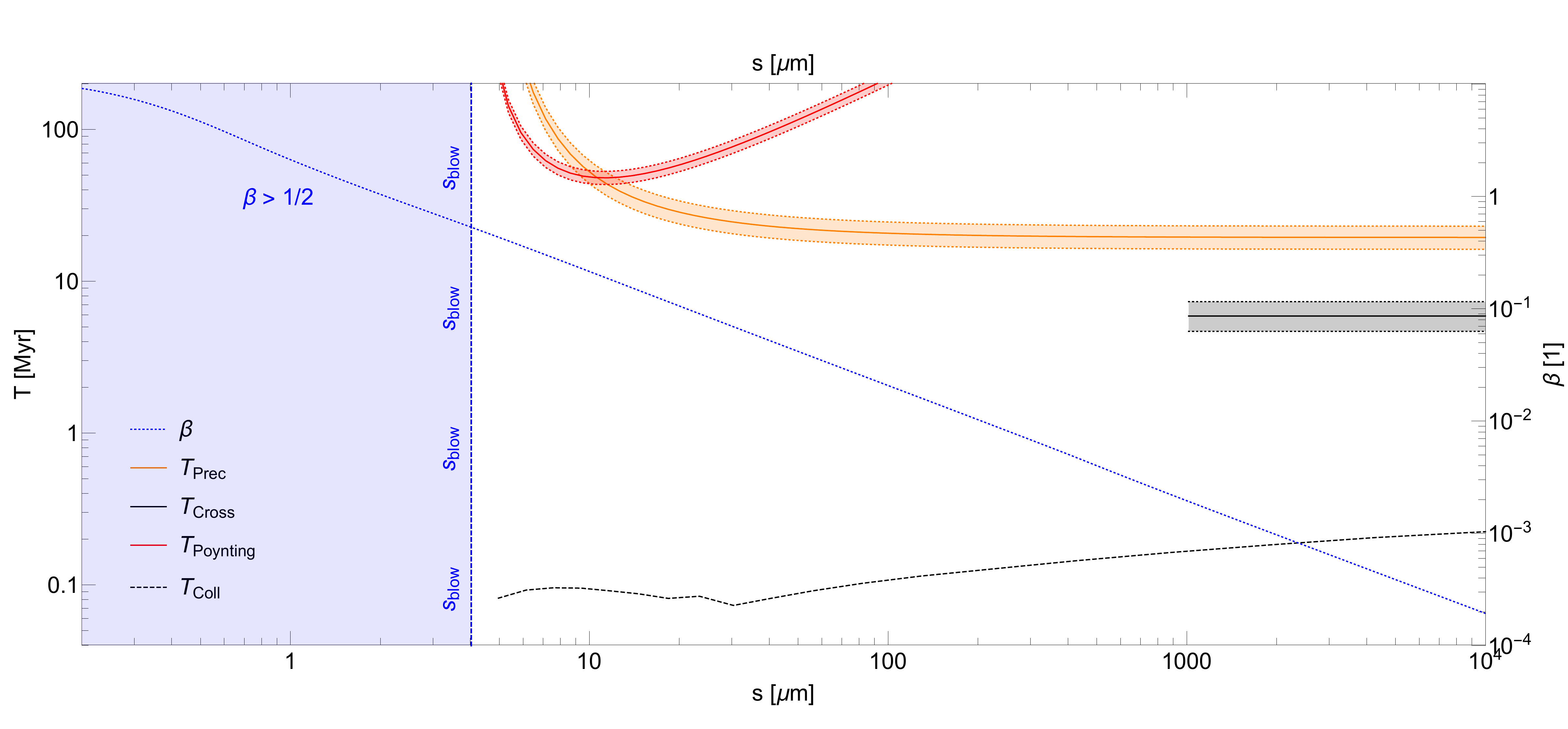}
	\caption{
		Relevant timescales for fragments created by collisions inside a debris disk ($a=95-105\mathrm{au}$; $m_\mathrm{Dust}=4.05\times 10^{25}\mathrm{g}$) around an A4V-star ($M_\star=1.92M_\odot$) with a planet at $a_\mathrm{p}=20\mathrm{au}$ with an eccentricity of $e_\mathrm{p}=0.6$ and a mass of $m_\mathrm{p}=2.5\mathrm{M_J}$. The $\beta$ value of the particles (\textit{blue, dotted}), and the blowout limit (\textit{blue zone on the left}) are shown. The precession timescale (\textit{orange; top}), the crossing timescale (\textit{black; middle}; where applicable), and the Poynting--Robertson timescale (\textit{red; crossing the top and the middle plot}) are derived analytically. The collision timescale (\textit{black, dashed}) was taken directly from the numerical results of the simulation. When possible, a range (\textit{shaded area}) of timescales for $a=95-105\mathrm{au}$ is given in addition to the value at $a=100\mathrm{au}$ (\textit{solid line}).
	}
	\label{fig:beta_tprec}
\end{figure*}

In our later simulation (Sect.~\ref{sec:implementation}), we are using an A4V-star. The relevant timescales are shown in Fig.~\ref{fig:beta_tprec}. The dust mass is $m_\mathrm{Dust}=4.05\times 10^{25}\mathrm{g}$ for particles smaller than $1\mathrm{mm}$, at a time of $5.75\mathrm{Myr}$. The blowout size is $s_\mathrm{blow}=4\mathrm{\mu m}$. For sizes greater than $100\mathrm{\mu m}$, $\beta$ is nearly zero ($\beta < 0.04$). These particles roughly follow the motion of the parent belt.\footnote{The graph is truncated at $10^4\mu m$, but even larger bodies exist in the simulation.} Because of the small $\beta$ value, the precession timescale is constant for a wide range of particle sizes. However, near the blowout size, it increases significantly with smaller particle size because these particles have larger semi-major axes. The Poynting--Robertson timescale decreases with smaller particle sizes because the influence of radiation pressure gets larger with smaller particle size. Only for the smallest grains, near the blowout limit, the timescale increases again because of the extremely eccentric, extended orbits the particles are on.

For the particle sizes shown, the collision timescale is always shorter than the precession timescale. Particles in this size range precess around the forced eccentricity in the $h$--$k$ space, but will collide with other particles before they can complete a total precession period. In contrast, the larger objects of the debris disk have much longer collisional lifetimes (for particles of $1\mathrm{m}$ already $T_\mathrm{Col}\ge 10\mathrm{Myr}$). They will complete one or more precession periods in the $h$--$k$ space before they are destroyed in collisions. That is, the larger debris in the parent belt precesses and simultaneously creates a halo of dust, which precesses more slowly, with the precession time depending on the $\beta$ value, and thus the particle size.

Fig.~\ref{fig:beta_tprec} shows the crossing timescale for particles with $\beta\approx 0$ (otherwise it does not apply). For these particles the crossing timescale is shorter than the precession timescale. Thus, large belt particles with $\beta\approx 0$ on non-crossing orbits with marginally different semi-major axis will evolve into a crossing configuration before they can complete a whole precession period, meaning that their collision rate increases. This effect might be significant and even trigger the collisional cascade \citep{mustill+wyatt2009}. Yet, in our later numerical example (see Sect.~\ref{sec:numerical_results}) we assume a pre-stirred disk, where it is less important.

The Poynting--Robertson timescale is mostly longer than the precession and crossing timescale, meaning that Poynting--Robertson drag does not play an important role. For particles smaller than $10\mathrm{\mu m}$, the Poynting--Robertson timescale is shorter than the precession timescale, meaning that these particles fall into the star due to Poynting--Robertson drag before they can complete a rotation in the $h$--$k$ space. Again, in our later example, this effect is not important because the collision timescale is even smaller.

The interplay between these timescales is dependent on the position of the planet $a_\mathrm{p}$ and the debris disk belt $a_{belt}$, in particular because the Poynting--Robertson and the collisional timescale only depend on the belt alone. For example, if the belt is placed at $50\mathrm{au}$ instead of $100\mathrm{au}$ (other parameters are the same as above), the timescales change as follows:
\begin{align}
	T_\mathrm{cross}(50\mathrm{au})&\approx\frac{1}{23}T_\mathrm{cross}(100\mathrm{au})\text{,}\nonumber\\
	T_\mathrm{PR}(50\mathrm{au})&\approx\frac{1}{4}T_\mathrm{PR}(100\mathrm{au})\text{,}\nonumber\\
	T_\mathrm{prec}^\mathrm{full}(50\mathrm{au})&\approx\frac{1}{11}T_\mathrm{prec}^\mathrm{full}(100\mathrm{au})\text{,}\nonumber\\
	T_\mathrm{col}(50\mathrm{au})&\approx\frac{1}{3}T_\mathrm{col}(100\mathrm{au})\footnote{The collisional timescales roughly scales with the orbital period.}\text{.}
\end{align}
All these timescales get smaller, however the ratio is much smaller for the crossing and the precession timescales, leading to much more pronounced differential precession. 

\section{Combining perturbations and collisions}\label{sec:implementation}
The model presented above provides estimates for individual pieces of the physics. Yet, for an analysis of the interplay between collisions and perturbations a numerical approach is needed. For that purpose we use our existing kinetic code \textit{ACE} \citep{krivov+2005,krivov+2006,reidemeister+2011,loehne+2017}. In \textit{ACE} the debris disk evolution is described by a four-dimensional phase space distribution. The dimensions are: the masses of the individual particles $m$, the pericentre distances $q$, the eccentricities $e$, and the longitudes of periapsis $\varpi$. We use the pericentre distance instead of the semi-major axis because the belt and the halo have similar pericentre distances, whereas the semi-major axes differ widely. In order to combine collisions with secular perturbations, we now add the gravitational influence of a single perturbing planet.

\subsection{Implementation in \textit{ACE}}
\textit{ACE} calculates secular perturbations using the transport equation. The phase space density $n(x)$ at the position $x = (m,q,e,\varpi)$ changes depending on the velocity $v(x)=\left[0,\dot{q}(x),\dot{e}(x),\dot{\varpi}(x)\right]$ at this point:
\begin{align}
	\label{eq:transport}
	\dot{n}(x)=&-\left[\frac{\partial}{\partial q}n(x)\dot{q}(x)+\frac{\partial}{\partial e}n(x)\dot{e}(x)+\frac{\partial}{\partial \varpi}n(x)\dot{\varpi}(x)\right]\text{.}
\end{align}
For each grid point $x$ in the four-dimensional phase space, the derivatives $\dot{e}(x)$ and $\dot{\varpi}(x)$ are calculated using the derivative of equation~(\ref{eq:hkdefinition}):
\begin{align}
	\label{eq:edotvarpidot}
	\dot{e}     =\dot{h}\sin{\varpi}+\dot{k}\cos{\varpi}\text{,} && \dot{\varpi}=\frac{1}{e}\left(\dot{h}\cos{\varpi}-\dot{k}\sin{\varpi}\right)\text{.}
\end{align}
The derivatives $\dot{h}(x)$ and $\dot{k}(x)$ are calculated using equation~(\ref{eq:hkdiff}), thus no additional analytical approximation is necessary. A series expansion of the Laplace coefficients is used.

The derivative of the pericentre distance needed in equation~(\ref{eq:transport}) is
\begin{align}
	\dot{q}=\dot{a}(1-e)-a\dot{e}\text{.}
\end{align}
The contributions to $\dot{e}$, $\dot{\varpi}$, and $\dot{a}$ are summed up for secular precession and Poynting--Robertson drag, with $\dot{a}=0$ for the former and $\dot{\varpi}=0$ for the latter. The transport equation itself is approximated using a variant of the upwind differencing scheme. \citep{press+2002} The parameter velocity $v(x)$ is calculated for each phase space bin, and the material shifted into the neighbouring bins depending on the velocity direction.

\begin{figure}
	\resizebox{\hsize}{!}{\includegraphics{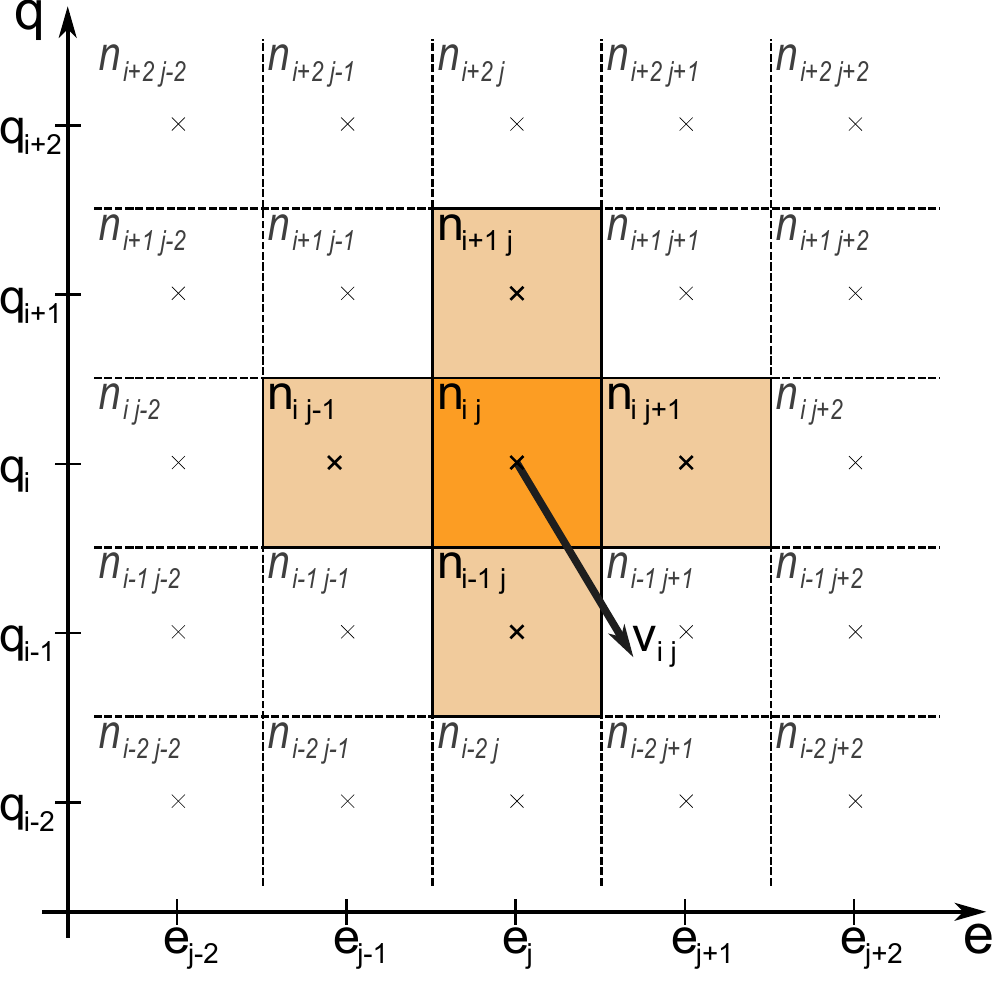}}
	\caption{
		Sketch of the \textit{upwind} scheme, used inside \textit{ACE}, acting on the pericentre $q$ and eccentricity $e$ dimensions of the four-dimensional phase space. The material is shifted from each bin $n_{i,j}$ into its direct neighbours by the parameter velocity $v_{i,j}$ (calculated with perturbation theory).
	}
	\label{fig:grid}
\end{figure}
A sketch of the algorithm is shown in Fig.~\ref{fig:grid}. \textit{ACE} uses a four-dimensional phase space, however for simplicity we only show a two-dimensional sketch of the pericentre $q$ and eccentricity $e$ dimensions. According to our differencing scheme, bin $n_{i,j}$ loses material to its neighbours and gains some from them. The amount lost depends on the parameter velocity $v_{i,j}$. The material lost in $n_{i,j}$ is moved into its direct neighbours along the $q$ and $e$ axis. In the example, these are the $n_{i,j+1}$ and the $n_{i-1,j}$ bin.

The total phase space change for the $n_{i,j}$ bin is:
\begin{align}
	\dot{n}_{i,j}&=v_{i-1,j}\mathbf{e}_q\frac{n_{i-1,j}}{\Delta q}\quad \text{(if $v_{i-1,j}\mathbf{e}_q > 0$)}\nonumber\\
	             &+v_{i+1,j}\mathbf{e}_q\frac{n_{i+1,j}}{\Delta q}\quad \text{(if $v_{i+1,j}\mathbf{e}_q < 0$)}\nonumber\\
				 &+v_{i,j-1}\mathbf{e}_e\frac{n_{i,j-1}}{\Delta e}\quad \text{(if $v_{i,j-1}\mathbf{e}_e > 0$)}\nonumber\\
				 &+v_{i,j+1}\mathbf{e}_e\frac{n_{i,j+1}}{\Delta e}\quad \text{(if $v_{i,j+1}\mathbf{e}_e < 0$)}\nonumber\\
				 &-v_{i,j}\mathbf{e}_e\frac{n_{i,j}}{\Delta e}-v_{i,j}\mathbf{e}_q\frac{n_{i,j}}{\Delta q}\text{,}
\end{align}
where $\Delta e$ and $\Delta q$ are the spacing of the phase space grid, and $\mathbf{e}_e$ and $\mathbf{e}_q$ are the unit vectors in $e$ and $q$ direction.

One shortcoming of this first-order method is that no material is moved directly along the direction that the parameter velocity $v_{i,j}$ is actually pointing to. In the example in Fig.~\ref{fig:grid}, no material moves directly into bin $n_{i-1,j+1}$. However, if we assume that the velocity is similar in the neighbouring bins, the discrepancy is compensated by material moving from bins $n_{i,j+1}$ and $n_{i-1,j}$ into $n_{i-1,j+1}$. Accuracy is limited by finite bin widths and parameter velocities varying within the grid.

The scheme used in \textit{ACE} introduces an artificial numerical dispersion into the calculation. Looking at the movement in only the $\varpi$ dimension, and assuming the parameter velocity to be constant, one gets the following differential equation:
\begin{align}
	\label{eq:dispersionone}
	\frac{\partial n\left(\varpi,t\right)}{\partial t}=-v\frac{\partial n\left(\varpi,t\right)}{\partial \varpi}\text{.}
\end{align}
Following the notation of \citet{press+2002}, and using our differentiation scheme, one gets:
\begin{align}
	\frac{\partial n\left(\varpi,t\right)}{\partial t}=-v\left(\frac{n^t_j-n^t_{j-1}}{\Delta\varpi}\right)\text{.}
\end{align}
This equation can be rearranged by adding $0 = n^t_{j+1}-n^t_{j+1}$:
\begin{align}
	\frac{\partial n\left(\varpi,t\right)}{\partial t}=-v\left(\frac{n^t_{j+1}-n^t_{j-1}}{2\Delta\varpi}\right)+\frac{v\Delta\varpi}{2}\left(\frac{n^t_{j+1}-2n^t_j+n^t_{j-1}}{\Delta\varpi^2}\right)\text{.}
\end{align}
Comparing the result with the one derived by \citet{press+2002} for the Forward-Time-Centre-Space scheme, one sees that the discretised equation is the same as the following differential equation:
\begin{align}
	\label{eq:dispersion}
	\frac{\partial n\left(\varpi,t\right)}{\partial t}=-v\frac{\partial n\left(\varpi,t\right)}{\partial \varpi}+D\frac{\partial^2 n\left(\varpi,t\right)}{\partial \varpi^2}
\end{align}
The equation is the same as equation~(\ref{eq:dispersionone}), but introduces an additional dispersion term with $D=v\Delta\varpi/2$. The dispersion term depends on the grid spacing $\Delta\varpi$ and the parameter velocity $v$, but is independent of the integration time step. Therefore, the initial phase space distribution always disperses over time.

\subsection{Collisionless test case}\label{sec:a_simple_implementation_test}
\begin{figure*}
	\centering
	\includegraphics[width=17cm]{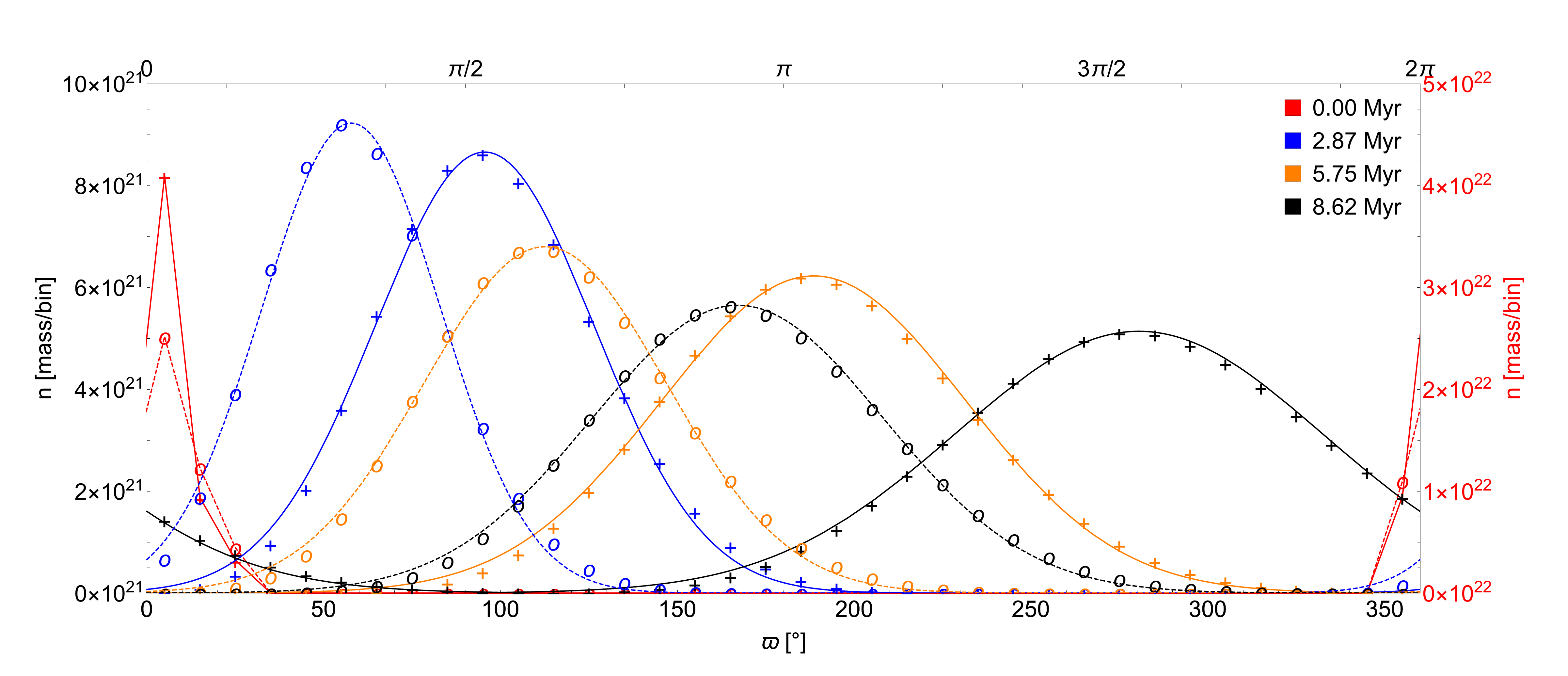}
	\caption{
		Azimuthal diffusion in the \textit{ACE} simulation described in Sec.~\ref{sec:a_simple_implementation_test}. The dust distribution of particles at $a=95\mathrm{au}$ (\textit{solid; crosses}) and $a=110\mathrm{au}$ (\textit{dashed; circles}) along the $\varpi$ dimension is plotted for a simulation time of $0.00\mathrm{Myr}$ (\textit{red}), $2.87\mathrm{Myr}$ (\textit{blue}), $5.75\mathrm{Myr}$ (orange), and $8.62\mathrm{Myr}$ (black). A fit to the analytical solution was made to extract the different parameter velocities and dispersion coefficients. The initial distribution is scaled down by a factor of $5$ to fit the plot.
	}
	\label{fig:implementation}
\end{figure*}

As an implementation check, we ran a simulation without collisions, but with an inner planet acting on the debris disk. The planet was set on a circular orbit ($e_\mathrm{p}=0$) at $a_\mathrm{p}=20\mathrm{au}$. The debris disk was set between $a=95\mathrm{au}$ and $110\mathrm{au}$, with the eccentricity being distributed in a radius of $0.1$ around $e=0.3$, $\varpi=0.0$ in the $h$--$k$ space. Inserting the data into equation~(\ref{eq:innerouter}), we see that the planet produces a precession of debris in the $h$--$k$ phase space around $e_\mathrm{forced}=0$. This precession does not modify the eccentricities of the orbits, only the longitudes of pericentre change over time. The precession frequency depends on the slight differences in semi-major axis and on the differences of the $\beta$ values of the grains. Without collisions, no new particles with different orbital configurations are created and equation~(\ref{eq:combination}) does not apply. The initial distribution is only moved around. Figure~\ref{fig:implementation} shows a $n$--$\varpi$ plot of the phase space distribution of objects with $s=52\mathrm{km}$ and a semi-major axis of $a=95\mathrm{au}$ and $110\mathrm{au}$. We chose the size of $s=52\mathrm{km}$ to make sure that the differences we see only depend on the semi-major axis and not on the $\beta$ value. For both distances the distribution moves over time, but because of the numerical dispersion explained in equation~(\ref{eq:dispersion}), it gets more spread out along $\varpi$ over time as well.

The simulated distribution was fitted to equation~(52) of \citet{neuman1981}. The fitted parameter velocities match the analytical expectations. For a semi-major axis of $a=110\mathrm{au}$, a parameter velocity of $v=17.94\mathrm{deg}/\mathrm{Myr}$ was predicted, and a velocity of $v=17.99\mathrm{deg}/\mathrm{Myr}$ was measured. For $a=95\mathrm{au}$ a velocity of $v=29.97\mathrm{deg}/\mathrm{Myr}$ was predicted and $v=31.10\mathrm{deg}/\mathrm{Myr}$ measured. The same was repeated for the dispersion factor $D$, using the estimate of $D=v\Delta\varpi/2$ from the previous section and the grid spacing of $\Delta\varpi=10\mathrm{deg}$. In the case of $a=110\mathrm{au}$, a dispersion of $D=89.7\mathrm{deg}^2/\mathrm{Myr}$ was predicted and a $D=95.3\mathrm{deg}^2/\mathrm{Myr}$ was measured. At $a=95\mathrm{au}$, $D=149.8\mathrm{deg}^2/\mathrm{Myr}$ was predicted and $D=152.1\mathrm{deg}^2/\mathrm{Myr}$ measured.

These results show that the simulation behaved as expected. The dispersion of the phase space distribution was exaggerated by additional numerical errors. The slight difference in numerical value of $v$ and $D$ were not further investigated, but are probably due to our usage of the \textit{upwind} scheme, which loses some accuracy in favour of fidelity \citep{press+2002}. The grid spacing of $\Delta\varpi=10\mathrm{deg}$ was kept for our later example as a compromise between accuracy and simulation runtime. For the eccentricity grid spacing, which was not important in the implementation test, the same rule holds. The phase space distribution disperses with $D=v\Delta e/2$ due to our integration scheme. However, the grid spacing is not linear in order to represent the orbits of particles created in collisions as well as possible. Therefore, the dispersion will depend on the position in the eccentricity grid. The dispersion could be reduced by adding more points to the grid, thereby reducing $\Delta\varpi$, or by making the now constant grid follow the perturbed motion through $e$--$\varpi$ space. Both approaches would significantly increase the simulation runtime.

Albeit numerical dispersion limiting the accuracy of the results, the main effects of secular perturbations combined with collisions remained visible in our simulation. The differential secular perturbations caused by differences in semi-major axes and $\beta$ values equalled or exceeded the numerical dispersion. At the given times, the $\varpi$ distributions at the inner and outer belt edges in Fig.~\ref{fig:implementation} were as far apart as they were wide, which can be considered as a minimum requirement. A comparison between particles with low and high $\beta$ values will yield larger differences in semi-major axis and therefore larger differences in precession velocity and larger separations in $\varpi$ as well.

\subsection{Model parameters}
We ran a full collisional simulation including perturbations by a single planet, and Poynting--Robertson drag. The planet orbited a A4V star, with a mass, luminosity, and effective temperature of $M=1.92M_\odot$, $L=16.6L_\odot$, and $T_\mathrm{eff}=8600\mathrm{K}$. The debris disk extended from $a=95\mathrm{au}$ to $105\mathrm{au}$, and had a moderate initial total mass of $m_\mathrm{Disk}=2M_\oplus$ (cf. \citet{krivov+2018}). In contrast to the test case in Sect.~\ref{sec:a_simple_implementation_test}, the debris was initially distributed on nearly circular orbits, spread between eccentricities $e=0$ and $e=0.1$. We assumed an initial mass distribution index of $q=1.888$, a minimum particle radius of $s_\mathrm{min}=100\mathrm{\mu m}$ and a maximum radius of $s_\mathrm{max}=49\mathrm{km}$. The perturbing planet had a mass, eccentricity and semi-major axis of $m_\mathrm{p}=2.5M_\mathrm{Jup}$, $e_\mathrm{p}=0.6$, $a_\mathrm{p}=20\mathrm{au}$, causing a forced eccentricity $e_\mathrm{forced} = 0.15$ at $a=100$~au.

The assumed material was a homogeneous mixture \citep{bruggeman1935} of Astrosilicate \citep{draine2003} and water ice \citep{li+greenberg1998} in equal volume fractions, with a bulk density of $2.35\mathrm{g}/\mathrm{cm}^3$. In order to calculate the $\beta$ values of the particles, and generate the emission images of the debris disk, the nearest point in the PHOENIX/NextGen grid was used \citep{hauschildt+1999}.

The eccentricity grid of the simulation span $[0.01,1.5]$ and was logarithmically spaced for $e\le 0.2$, and for $1 < e$, with at most a factor of $\Delta e/e\approx0.5$ between neighbouring bins. For $0.2 < e < 1$, the step size was limited to $\Delta e\le 0.1$. The transition between the linear and logarithmic regimes are smooth. For a more detailed description of the grid generation, see eq.~(29) in \citet{loehne+2017}. The orientation grid had a constant spacing of $\Delta\varpi=10\mathrm{deg}$. The pericentre grid had $60$ logarithmically spaced grid points, spanning from $40\mathrm{au}$ to $160\mathrm{au}$. The mass grid span a range from $7.14\times 10^{-13}\mathrm{g}$ to $1.13\times 10^{21}\mathrm{g}$. Objects unaffected by radiation pressure had a logarithmic spacing of $\Delta m/m\approx 12$. Small dust grains had a logarithmic spacing of $\Delta m/m\approx 3$. The transition between the two regimes is smooth, modelled with the number of bins per unit logarithm of mass being
\begin{align}
	\frac{\mathrm{d}j}{\mathrm{d}\ln{m}}=\frac{1}{12}\left[1+3\left(1-e^{-20\beta}\right)\left(\frac{1}{2}+\frac{\tan^{-1}\left[3\left(2-\beta\right)\right]}{\pi}\right)\right]\text{.}
\end{align}

\subsection{Numerical Results}\label{sec:numerical_results}
\begin{figure*}
	\centering
	\includegraphics[width=17cm]{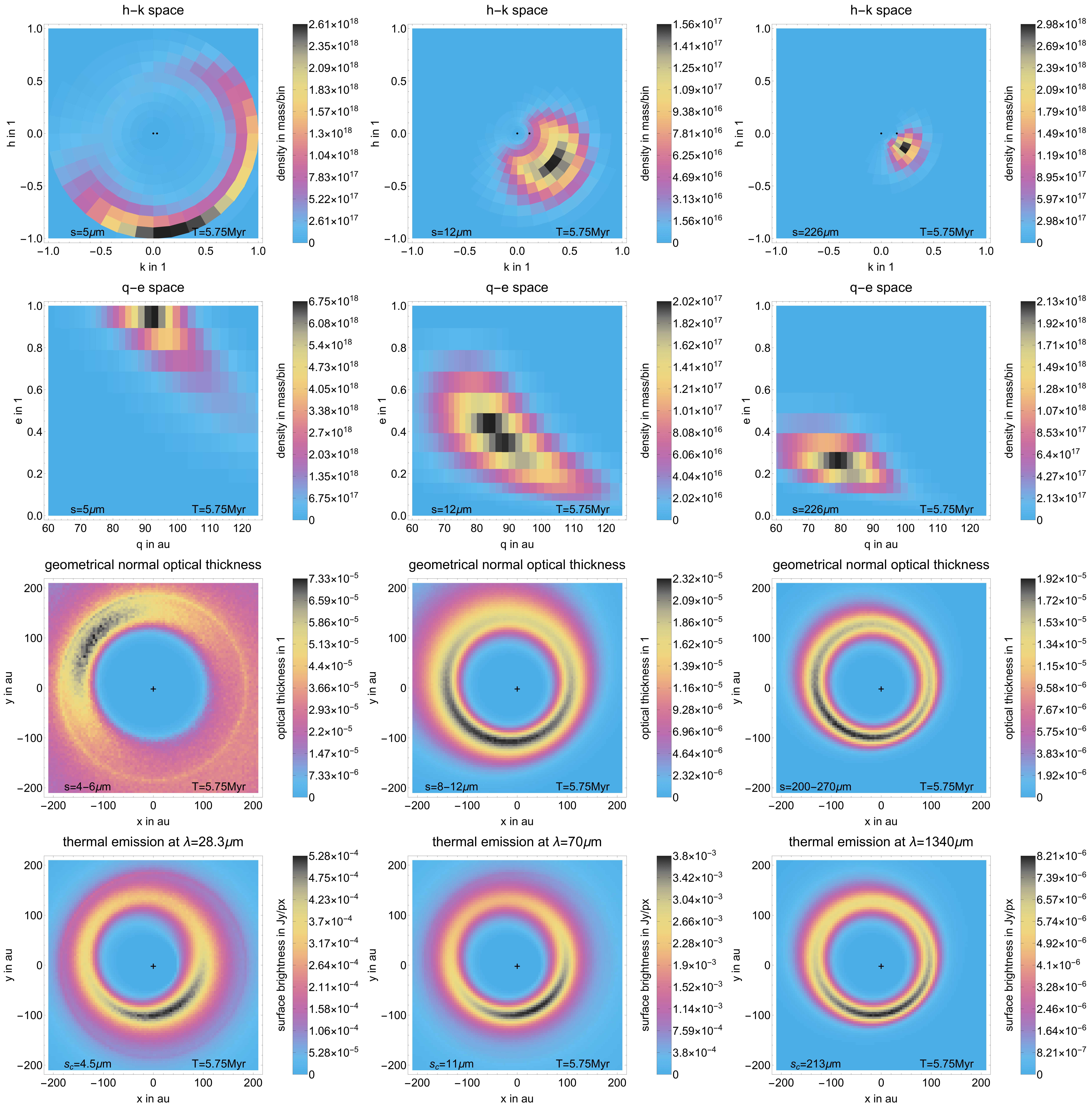}
	\caption{
		Results of the \textit{ACE} simulation described in Sec.~\ref{sec:numerical_results}. The \textit{first row} shows the $h$--$k$ phase space distribution, the \textit{second row} shows the $q$--$e$ distribution, and the \textit{third row} shows the corresponding geometrical optical thickness. Each column shows a different particle size range: $\approx 5\mathrm{\mu m}$ ($\beta\approx 0.43$) on the \textit{left}, $\approx 12\mathrm{\mu m}$ ($\beta\approx 0.18$) in the \textit{middle}, and $\approx 226\mathrm{\mu m}$ ($\beta\approx 0.009$) on the \textit{right}. The \textit{fourth row} shows the thermal emission over all particle sizes (without applying any instrument function). The characteristic wavelengths shown, match the particle sizes from the plots above.
	}
	\label{fig:phase_space_evolution}
\end{figure*}

Figure~\ref{fig:phase_space_evolution} shows how the initial distribution has evolved after $T=5.75\mathrm{Myr}$. The two top rows show the $h$--$k$ space and the $q$--$e$ space for grains in bins centred around radii of $\approx 5\mathrm{\mu m}$, $\approx 12\mathrm{\mu m}$, and $\approx 226\mathrm{\mu m}$. (Single mass bins encompassing a size range are shown.) In the initial distribution grains down to $s_\mathrm{min}=100\mathrm{\mu m}$ were generated. Grains of size $\approx 5\mathrm{\mu m}$ and $\approx 12\mathrm{\mu m}$ are not part of the initial distribution, but were created by collisions within the disk. Particles of size $\approx 226\mathrm{\mu m}$ have been initially present, but as shown in Fig.~\ref{fig:beta_tprec}, their collisional lifetime is very short. They did not survive until the simulation time of $T=5.75\mathrm{Myr}$, but were replenished by collisions of larger bodies.

Grains of size $5\mathrm{\mu m}$, $12\mathrm{\mu m}$, and $226\mathrm{\mu m}$ have $\beta$ values of $0.43$, $0.18$, and $0.009$. According to equation~(\ref{eq:precscale}) the precession timescales are $1990\mathrm{Myr}$, $42\mathrm{Myr}$, and $19\mathrm{Myr}$ (at a distance of $a_\mathrm{belt}=100\mathrm{au}$). Thus it is clear, that at the current simulation time of $T=5.75\mathrm{Myr}$ no debris in the disk has completely precessed around the forced eccentricity. It is still the initial stage of precession.

The top row of Fig.~\ref{fig:phase_space_evolution} matches the theory of differential secular perturbations. Grains of all size ranges precess counter-clockwise around the forced eccentricity. The $\approx 5\mathrm{\mu m}$ and $\approx 12\mathrm{\mu m}$ grains show precession differential to the $\approx 226\mathrm{\mu m}$ grains, which follow the movement of the parent belt. Due to their large $\beta$ values, small grains are distributed in a larger area of the phase space.

The distribution of $\approx 5\mathrm{\mu m}$ grains seems counter intuitive. According to our theoretical predictions, grains should be distributed in a large circle at $e\lesssim 1$. However, this is only true under the assumption of collisions in a circular belt. Barely bound grains produced by an already elliptic belt are not uniformly distributed in $\varpi$. Collisions at the pericentre of the disk are more likely to produce unbound grains than collisions at the apocentre, explaining the excess of particles in the bottom--right. Furthermore, bins near $e=1$ are wider, containing more mass per bin, which results in an apparent density increase at the bottom--right. (See Fig.~3 in \citet{loehne+2017} as well.)

In addition, for some of the $\approx 5\mathrm{\mu m}$ grains, the analytic prediction suggests that they should reach eccentricities of $e>1$. This is a shortcoming of the linear approximation of the Laplace--Lagrange theory. Secular perturbations cannot make a particle cross the $e=1$ boundary because bringing particles on unbound orbits would require energy transfer. (Which is forbidden by $\dot{a}=0$.) In \textit{ACE} simulations this is handled by setting $\dot e \le 0$ for the bin closest to $e=1$, while still allowing arbitrary $\dot\varpi$. This handling of the boundaries reduces the fidelity of the results for very eccentric grains. However, the phase space asymmetry with respect to the line running from the top--left to the bottom--right is no pure artefact as these grains have large semi-major axes, long precession timescales and lifetimes, and hence, move slowly through the phase space, allowing pile-up near $e = 1$. A more detailed analysis would require a finer eccentricity grid in this region of the phase space.

The phase space evolution discussed in Sect.~\ref{sec:differential_precession} and Sect.~\ref{sec:timescales} does not take into account that the initial grains are destroyed over time. Yet, although these particles are destroyed in collisions, new particles are created by other collisions as well. At any moment in time, a new starting distribution of particles is created, which shows differential precession over time. An image of a disk is always a superposition of different, initial distributions. As these initial distributions evolve over time, they occupy different parts of the $h$--$k$ space, effectively looking like a single, smeared out distribution. At the same time, dust of older initial distributions is lost in collisions, making them fade in intensity. Therefore observations always show a smeared out snapshot of the most recent initial distributions. In order to extract the correct planet parameters from an image, one needs to simulate the whole phase space evolution to get the smearing right. However, in an approximation, the model described in Sect.~\ref{sec:differential_precession} and Sect.~\ref{sec:timescales} can still be used.

The second row of Fig.~\ref{fig:phase_space_evolution} shows the $q$--$e$ distribution for the same particle sizes. As explained by equation~(\ref{eq:collision}), particles with a lower $\beta$ are created on more eccentric orbits, therefore the smaller particles have higher eccentricities. However, the eccentricity is additionally increased by the secular perturbations of the planet, leading to higher eccentricities even for the parent belt: most $\approx 226\mathrm{\mu m}$ grains have reached $e>0.2$, instead of staying at the initial $e=0$ to $e=0.1$. This change in eccentricity is reflected in the pericentre distribution as well. With the belts semi-major axis starting between $95\mathrm{au}$ and $105\mathrm{au}$, the pericentres have now evolved to approximately $76\mathrm{au}$ to $84\mathrm{au}$.

The same is true for the numerical diffusion. Due to the coupling of pericentre distance and eccentricity, a diffusion in eccentricity leads to a diffusion in pericentre distance as well. This limits the simulation time, as for long enough timescales, the differences in semi-major axis, which cause differential precession, are lost in the diffusion of pericentre distances.

The third row shows the geometrical normal optical thickness (hereafter referred to as optical thickness) of the disk for certain grain size ranges. The optical thickness is the amount of material in the sight line, and while not observable directly, it provides a good illustration of the spatial distribution of material in the disk.

Except for barely bound grains ($s=4-6\mu\mathrm{m}$), the optical thickness maps roughly match the phase space distributions of the top two rows. The orientation of the disk slightly differs between different particle sizes, as expected for the differences in orientation shown in the $h$--$k$ plots. Furthermore, in the maps for both $\approx 12\mathrm{\mu m}$ and $\approx 226\mathrm{\mu m}$, we observe the following: the disk is visibly offset from the star; the disk is slightly eccentric; the material density in the disk is the highest south of the star, and not at the closest distance to it. Moreover, the disks get larger for smaller grain sizes.

The offset and the eccentricity of the disk are explained by the planet shaping it. It is more extended for smaller particle sizes because, due to their $\beta$ value, these particles are on more eccentric orbits with longer semi-major axes. The asymmetry in particle density is a geometric effect. An example of which can be seen in Fig.~\ref{fig:idea}. On the south side, the particle orbits are arranged such that the paths of multiple orbits overlap nearly perfectly, resulting in a higher density. On the other sides, the paths of the orbits diverge significantly, resulting in a lower density. The same effect is discussed in detail by \citet{wyatt2005b}.

The optical thickness of barely bound grains ($s=4-6\mu\mathrm{m}$) shows a different behaviour. It is higher at the apocentre because in collisions of parent bodies at the pericentre, the $\beta$ values are high enough for these small fragments to become unbound. For collisions at the apocentre, the fragments are still bound. This effect is also visible in Fig.~9 in \citet{loehne+2017} in logarithmic scale. Because this dependence on the collision point restricts the orbits (in addition to the secular perturbations), the axial asymmetry gets diminished.

The fourth row shows the linearly scaled thermal emission surface brightness of the debris disk at $\lambda=28.3\mathrm{\mu m}$, $70\mathrm{\mu m}$ and $1340\mathrm{\mu m}$. The wavelengths were chosen to closely match astronomical telescopes, and for their associated characteristic grain size to match the particle sizes chosen for the other rows. In the presented case, $70\mathrm{\mu m}$ observations could have been made by Herschel/PACS, the $1340\mathrm{\mu m}$ observation could be made by ALMA, and the $28.3\mu\mathrm{m}$ observations could be made with JWST.

Grains of all sizes contribute to thermal emission and are taken into account in our modelling. Yet, at a given wavelength $\lambda$, the emission is dominated by a limited range of grain sizes around the characteristic grain size $s_c=\lambda/\left(2\pi\right)$. \citep{backman+paresce1993} In that sense, the wavelengths of the thermal emission images match the grain sizes shown in the phase space plots. We therefore expect the images to roughly reflect the features and differences of the above phase space distributions, albeit with lower significance because of the convolution over a size range. An example is the shortest wavelength ($28.3\mu\mathrm{m}$), where the peculiarities of the smallest, barely bound grains ($s=4-6\mu\mathrm{m}$) are not reflected.

All thermal emission images show significant asymmetries. For all wavelengths we observe the following: the disk is visibly offset from the star; the disk is slightly eccentric; the parts of the disk closer to the star are brighter; the parts of increased brightness are not symmetric to the axis between the closest distance of the disk to the star and the star itself.

Again, the observed offset of the disk and its eccentricity is due to the planet shaping the initially circular disk into an ellipse. The disk is brighter at the pericentre\footnote{Contrary to \citet{wyatt2005b} and \citet{pan+2016} we do not observe apocentre glow because our images are not PSF broadened. See Fig.~(15) of \citet{loehne+2017} for the effect of PSF broadening.}, despite having less material than other regions, because the thermal emission intensity depends on the temperature of the grains. Particles spend more time at the apocentre of their orbit, where they cool down and do not contribute as much to thermal emission, as for the short period they spend close to the star, where they are heated up. The thermal emission therefore depends on the distance to the star. This effect is more pronounced for shorter wavelengths due to the exponential temperature dependence in the Wien regime, leading to a stronger radial dependence \citep{wyatt+1999}.

The axial asymmetry in the thermal emission derives from the optical thickness distribution, which shows more material south of the star. However, due to the pericentre glow the asymmetry is not as strong as for the optical thickness.

\section{Discussion}\label{sec:discussion}
\subsection{Relevance of differential precession}
The crucial difference between the simulations shown in \citet{loehne+2017} and the models presented here is how the actual secular perturbations are implemented. Previously, we started with pre-perturbed belts that had already acquired eccentricity. The disks subsequently evolved only through the collisional cascade and drag forces, without ongoing dynamical perturbation. Now, secular perturbation and the resulting eccentricity evolution are modelled together.

Because \citet{loehne+2017} did not account for ongoing, differential precession, their results are applicable only if (a) the perturbations have stopped or (b) collisional replenishing dominates the evolution. Condition (a) could be fulfilled in a multi planet system, where a planet shapes the disk into an ellipse, but is then kicked out of the system in a close encounter with a more massive planet or sub-stellar companion closer to the star. The belt would still be precessing, albeit at a lower rate and with a forced eccentricity closer to zero. If the eccentric belt itself is massive enough \citep[cf.][]{krivov+2018}, it will however induce self-precession and differential precession of its halo. 
Condition (b) is met if, for example, the mass of the perturbing planet is low and the resulting precession timescales much longer than collision timescales. Under such circumstances the perturber would still make the belt elliptic, but differential precession could not accumulate because particles would be too short-lived to significantly trail (or lead) freshly produced ones.

Vice versa, differential precession can be significant if precession timescales are sufficiently short because either the perturber or an already eccentric belt are massive enough.

\subsection{Multiple perturbers}
Our main reason for presenting the case of a single perturbing planet is simplicity. While many multi-planet systems have been observed, the influence of a single, massive planet is much easier to implement and describe because the planet's orbital parameter do not evolve over time. The basic effects of differential precession can thus be shown more clearly. Our results do thus cover systems where one planet has a dominant influence on the disk, either because of proximity or because of mass.

It is possible to describe secular perturbation for multiple perturbers using the same analytic approximations in the $h$--$k$ space \citep{murray+dermott2000} and an according implementation in ACE. As a result of the separate and the mutual perturbations of the planets, more complex patterns for precession and differential precession would arise. Given an observed snapshot of an eccentric belt, the already present degeneracy among potential perturber parameters would increase. However, deeper analysis is beyond the scope of this paper.

\subsection{Finding hidden planets}
If high resolution images at several wavelengths are available, the effects of differential precession on the size distribution can be spatially resolved. Ideally, the wavelengths and corresponding characteristic grain sizes would cover $\beta(s)$ values from $\beta \approx 0.5$ to $\beta \approx 0$, that is, from barely bound grains in the halo to large grains that stay in the parent belt. Then, for example, the strength of the misalignment between spots of maximum brightness, as discussed in Sect.~\ref{sec:numerical_results}, can be used to derive timescales of differential precession relative to the collision timescales for grains in the corresponding size range. If collision timescales are estimated from the relative abundance of halo grains with respect to the belt \citep{thebault+wu2008}, the absolute timescales for differential precession can be estimated. Planetary mass and semi-major axis follow from equation~(\ref{eq:dispersion}).

In a collisionally active disk such as the one shown in Fig.~\ref{fig:phase_space_evolution}, only the larger grains, and with them the parent belt itself, will have collision timescales long enough for significant differential precession to occur. However, for these large grains the spread in $\beta$ values is necessarily narrow and size segregation will be unimportant. Precession will then be differential mainly with respect to semi-major axis, with the belt edge that is closer to the perturber precessing at a higher rate.

Given that the dust nearest to the star is brightest and that detection is most significant in bright regions of a disk, the location of the inner disk edge could in principle be measured as a function of observation wavelength. The inner edge is defined by the pericentre distances of the grains of corresponding characteristic size. Assuming that the belt has not evolved far from circularity, we approximate the evolution of the orbital eccentricities and the resulting pericentre distances from equation~(\ref{eq:combination}) as a function of time $t$:
\begin{align}
	\label{eq:findplanet}
	q_\mathrm{min}=a_\mathrm{belt}\left(1-\frac{2\pi}{T_\mathrm{belt}^\mathrm{full}}\ t\ e_\mathrm{belt}^\mathrm{forced}\left[\frac{1-\beta}{1-2\beta}\right]^{\mp 7/2}\right),
\end{align}
where the minus in the exponent is for an inner, the plus for an outer perturber. The precession timescale $T_\mathrm{belt}^\mathrm{full}$ and the forced eccentricity $e_\mathrm{belt}^\mathrm{forced}$ of the belt are calculated using equation~(\ref{eq:combination}) with $\beta=0$ and $a=a_\text{belt}$.

Equation~(\ref{eq:findplanet}) has effectively two free parameters, the planet's semi-major axis and eccentricity. Thus, fitting it to observations cannot untangle the planet parameters completely. Competing mechanisms, such as chaotic motion in the zone of overlapping mean-motion resonances near a perturber \citep{mustill+wyatt2012}, further complicate the analysis.

In principle, equation~(\ref{eq:findplanet}) could provide a simple constraint for possible planets in a debris disk system. However, as has been shown in this work, its practical application is limited by images being convolutions of grain size and orbit distributions. In contrast to thermal emission, scattered light images, which are more dominated by the isotropically scattering small grains, might retain more of the asymmetries. However, in that case, analysis is more sensitive to viewing geometry and grain properties. A more detailed discussion of scattered light is beyond the scope of this work.

\subsection{Other approaches to differential precession}
The modelling process presented above -- dynamics implemented as advection terms in a grid-based collisional code -- is computationally expensive. Analytic solutions would be desirable. However, such solutions are usually confined to the purely dynamical case, which is a good approximation only to the big objects with long collision lifetimes. For example, \citet{mustill+wyatt2009} discuss the timescale for differential precession by a planet to induce orbit crossing in the belt, thus stirring the belt and triggering the collisional cascade. The actual subsequent collisional evolution cannot be modelled purely analytically.

Another possible modelling approach is to start from an \textit{N}-body code that can handle dynamics and add the generation and removal of particles in collisions. SMACK \citep{nesvold+2013} and LIDT-DD \citep{kral+2013} are two examples that follow this approach. LIPAD \citep{levison+2012} is a third example, albeit one less focused on dust. In such models, radiation pressure and gravitational interaction with as many perturbers as desired is easily implemented. Short-term and resonant perturbation or the aftermaths of single, giant collisions can be followed. While these short-term phenomena are the domain of the dynamics-based models, we consider the orbit-averaged statistical approach of ACE to be better suited for smooth long-term evolution and secular perturbations. A compromise between the two approaches could be based on variable instead of the fixed grid points used in ACE. Secular perturbation would then move the grid points along with the material instead of redistributing material in the grid. Compared to our current implementation in ACE this would eliminate the numerical dispersion discussed in Sect.~\ref{sec:a_simple_implementation_test}. As a downside, the moving grid points would require collision properties to be evaluated again for each time step.

\section{Summary}\label{sec:summary}
The evolution of debris disks under the combined action of a collisional cascade and secular gravitational perturbations were simulated fully consistently for the first time with our code \textit{ACE}. We came to the following main conclusions:
\begin{enumerate}
	\item The timescales on which secular perturbations create elliptic rings from initially circular distributions are size-dependent through the dust grains' radiation pressure-to-gravity ratio $\beta$.
	\item Differential precession, if observed, can help disentangle a perturber's parameters, such as its mass and semi-major axis.
	\item When collision lifetimes are long enough, differential precession can notably shear the small-grain halo with respect to the larger objects in the parent belt. If collision lifetimes are short, shearing due to differential precession can only be observed within the belt itself.
	\item The detection of planets with this method is pivoting on the availability of images at different wavelengths, covering a wide range of grains sizes, from barely bound halo grains to big debris unaffected by radiation pressure.
\end{enumerate}

%Acknowledgements
\begin{acknowledgements}
We thank the anonymous referee for constructive remarks that helped clarify critical sections of the manuscript.

We are grateful for support by the Deutsche Forschungsgemeinschaft (project Lo 1715/2-1, Research Unit FOR\,2285 ``Debris Disks in Planetary Systems'').
\end{acknowledgements}
%References
\bibliographystyle{aa}
\bibliography{bibliography}

\begin{thebibliography}{48}
\expandafter\ifx\csname natexlab\endcsname\relax\def\natexlab#1{#1}\fi

\bibitem[{Backman \& Paresce(1993)}]{backman+paresce1993}
Backman, D.~E. \& Paresce, F. 1993, Main-sequence stars with circumstellar
  solid material - The VEGA phenomenon (University of Arizona Press),
  1253--1304

\bibitem[{Boley {et~al.}(2012)Boley, Payne, Corder, Dent, Ford, \&
  Shabram}]{boley+2012}
Boley, A.~C., Payne, M.~J., Corder, S., {et~al.} 2012, \apjl, 750, L21

\bibitem[{Bruggeman(1935)}]{bruggeman1935}
Bruggeman, D.~A.~G. 1935, Annalen der Physik, 416, 636

\bibitem[{Burns {et~al.}(1979)Burns, Lamy, \& Soter}]{burns+1979}
Burns, J.~A., Lamy, P.~L., \& Soter, S. 1979, Icarus, 40, 1

\bibitem[{Draine(2003)}]{draine2003}
Draine, B.~T. 2003, \apj, 598, 1017

\bibitem[{Eiroa {et~al.}(2013)Eiroa, Marshall, Mora, Montesinos, Absil,
  Augereau, Bayo, Bryden, Danchi, del Burgo, Ertel, Fridlund, Heras, Krivov,
  Launhardt, Liseau, Löhne, Maldonado, Pilbratt, Roberge, Rodmann,
  Sanz-Forcada, Solano, Stapelfeldt, Thébault, Wolf, Ardila, Arévalo,
  Beichmann, Faramaz, González-García, Gutiérrez, Lebreton,
  Martínez-Arnáiz, Meeus, Montes, Olofsson, Su, White, Barrado, Fukagawa,
  Grün, Kamp, Lorente, Morbidelli, Müller, Mutschke, Nakagawa, Ribas, \&
  Walker}]{eiroa+2013}
Eiroa, C., Marshall, J.~P., Mora, A., {et~al.} 2013, \aap, 555, A11

\bibitem[{Hauschildt {et~al.}(1999)Hauschildt, Allard, \&
  Baron}]{hauschildt+1999}
Hauschildt, P.~H., Allard, F., \& Baron, E. 1999, \apj, 512, 377

\bibitem[{Kalas {et~al.}(2005)Kalas, Graham, \& Clampin}]{kalas+2005}
Kalas, P., Graham, J.~R., \& Clampin, M. 2005, \nat, 435, 1067

\bibitem[{Kennedy {et~al.}(2018)Kennedy, Marino, Matrà, Panić, Wilner, Wyatt,
  \& Yelverton}]{kennedy+2018a}
Kennedy, G.~M., Marino, S., Matrà, L., {et~al.} 2018, \mnras, 475, 4924

\bibitem[{Kim {et~al.}(2018)Kim, Wolf, Löhne, Kirchschlager, \&
  Krivov}]{kim+2018}
Kim, M., Wolf, S., Löhne, T., Kirchschlager, F., \& Krivov, A.~V. 2018, \aap,
  618, A38

\bibitem[{Kral {et~al.}(2013)Kral, Thébault, \& Charnoz}]{kral+2013}
Kral, Q., Thébault, P., \& Charnoz, S. 2013, \aap, 558, A121

\bibitem[{Krivov {et~al.}(2018)Krivov, Ide, Löhne, Johansen, \&
  Blum}]{krivov+2018}
Krivov, A.~V., Ide, A., Löhne, T., Johansen, A., \& Blum, J. 2018, \mnras,
  474, 2564

\bibitem[{Krivov {et~al.}(2006)Krivov, Löhne, \& Sremčević}]{krivov+2006}
Krivov, A.~V., Löhne, T., \& Sremčević, M. 2006, \aap, 455, 509

\bibitem[{Krivov {et~al.}(2005)Krivov, Sremčević, \& Spahn}]{krivov+2005}
Krivov, A.~V., Sremčević, M., \& Spahn, F. 2005, \icarus, 174, 105

\bibitem[{Kuchner \& Holman(2003)}]{kuchner+holman2003}
Kuchner, M.~J. \& Holman, M.~J. 2003, \apj, 588, 1110

\bibitem[{Lee \& Chiang(2016)}]{lee+chiang2016}
Lee, E.~J. \& Chiang, E. 2016, \apj, 827, 125

\bibitem[{Levison {et~al.}(2012)Levison, Duncan, \& Thommes}]{levison+2012}
Levison, H.~F., Duncan, M.~J., \& Thommes, E. 2012, \aj, 144, 119

\bibitem[{Löhne {et~al.}(2017)Löhne, Krivov, Kirchschlager, Sende, \&
  Wolf}]{loehne+2017}
Löhne, T., Krivov, A.~V., Kirchschlager, F., Sende, J.~A., \& Wolf, S. 2017,
  \aap, 605, A7

\bibitem[{Löhne {et~al.}(2008)Löhne, Krivov, \& Rodmann}]{loehne+2008}
Löhne, T., Krivov, A.~V., \& Rodmann, J. 2008, \apj, 673, 1123

\bibitem[{Li \& Greenberg(1998)}]{li+greenberg1998}
Li, A. \& Greenberg, J.~M. 1998, \aap, 331, 291

\bibitem[{MacGregor {et~al.}(2017)MacGregor, Matrà, Kalas, Wilner, Pan,
  Kennedy, Wyatt, Duchene, Hughes, Rieke, Clampin, Fitzgerald, Graham, Holland,
  Panić, Shannon, \& Su}]{macgregor+2017}
MacGregor, M.~A., Matrà, L., Kalas, P., {et~al.} 2017, \apj, 842, 8

\bibitem[{Maldonado {et~al.}(2012)Maldonado, Eiroa, Villaver, Montesinos, \&
  Mora}]{maldonado+2012}
Maldonado, J., Eiroa, C., Villaver, E., Montesinos, B., \& Mora, A. 2012, \aap,
  541, A40

\bibitem[{Matrà {et~al.}(2018)Matrà, Marino, Kennedy, Wyatt, Öberg, \&
  Wilner}]{matra+2018}
Matrà, L., Marino, S., Kennedy, G.~M., {et~al.} 2018, ArXiv e-prints
  [\eprint{1804.01094}]

\bibitem[{Moerchen {et~al.}(2011)Moerchen, Churcher, Telesco, Wyatt, Fisher, \&
  Packham}]{moerchen+2011}
Moerchen, M.~M., Churcher, L.~J., Telesco, C.~M., {et~al.} 2011, \aap, 526, A34

\bibitem[{Montesinos {et~al.}(2016)Montesinos, Eiroa, Krivov, Marshall,
  Pilbratt, Liseau, Mora, Maldonado, Wolf, Ertel, Bayo, Augereau, Heras,
  Fridlund, Danchi, Solano, Kirchschlager, del Burgo, \&
  Montes}]{montesinos+2016}
Montesinos, B., Eiroa, C., Krivov, A.~V., {et~al.} 2016, \aap, 593, A51

\bibitem[{Murray \& Dermott(2000)}]{murray+dermott2000}
Murray, C.~D. \& Dermott, S.~F. 2000, Solar System Dynamics (Cambridge
  University Press)

\bibitem[{Mustill \& Wyatt(2009)}]{mustill+wyatt2009}
Mustill, A.~J. \& Wyatt, M.~C. 2009, \mnras, 399, 1403

\bibitem[{Mustill \& Wyatt(2012)}]{mustill+wyatt2012}
Mustill, A.~J. \& Wyatt, M.~C. 2012, \mnras, 419, 3074

\bibitem[{Nesvold {et~al.}(2013)Nesvold, Kuchner, Rein, \& Pan}]{nesvold+2013}
Nesvold, E.~R., Kuchner, M.~J., Rein, H., \& Pan, M. 2013, \apj, 777, 144

\bibitem[{Neuman(1981)}]{neuman1981}
Neuman, S.~P. 1981, Journal of Computational Physics, 41, 270

\bibitem[{Ozernoy {et~al.}(2000)Ozernoy, Gorkavyi, Mather, \&
  Taidakova}]{ozernoy+2000}
Ozernoy, L.~M., Gorkavyi, N.~N., Mather, J.~C., \& Taidakova, T.~A. 2000,
  \apjl, 537, L147

\bibitem[{Pan {et~al.}(2016)Pan, Nesvold, \& Kuchner}]{pan+2016}
Pan, M., Nesvold, E.~R., \& Kuchner, M.~J. 2016, \apj, 832, 81

\bibitem[{Pawellek {et~al.}(2014)Pawellek, Krivov, Marshall, Montesinos,
  Ábrahám, Moór, Bryden, \& Eiroa}]{pawellek+2014}
Pawellek, N., Krivov, A.~V., Marshall, J.~P., {et~al.} 2014, \apj, 792, 65

\bibitem[{Press {et~al.}(2007)Press, Teukolsky, Vetterling, \&
  Flannery}]{press+2002}
Press, W.~H., Teukolsky, S.~A., Vetterling, W.~T., \& Flannery, B.~P. 2007,
  Numerical Recipes: The Art of Scientific Computing (Cambridge University
  Press)

\bibitem[{Reidemeister {et~al.}(2011)Reidemeister, Krivov, Stark, Augereau,
  Löhne, \& Müller}]{reidemeister+2011}
Reidemeister, M., Krivov, A.~V., Stark, C.~C., {et~al.} 2011, \aap, 527, A57

\bibitem[{Rhee {et~al.}(2007)Rhee, Song, Zuckerman, \& McElwain}]{rhee+2007}
Rhee, J.~H., Song, I., Zuckerman, B., \& McElwain, M. 2007, \apj, 660, 1556

\bibitem[{Rodigas {et~al.}(2014)Rodigas, Malhotra, \& Hinz}]{rodigas+2014}
Rodigas, T.~J., Malhotra, R., \& Hinz, P.~M. 2014, \apj, 780, 65

\bibitem[{Schneider {et~al.}(2016)Schneider, Grady, Stark, Gaspar, Carson,
  Debes, Henning, Hines, Jang-Condell, Kuchner, Perrin, Rodigas, Tamura, \&
  Wisniewski}]{schneider+2016}
Schneider, G., Grady, C.~A., Stark, C.~C., {et~al.} 2016, \aj, 152, 64

\bibitem[{Schneider {et~al.}(1999)Schneider, Smith, Becklin, Koerner, Meier,
  Hines, Lowrance, Terrile, Thompson, \& Rieke}]{schneider+1999}
Schneider, G., Smith, B.~A., Becklin, E.~E., {et~al.} 1999, \apjl, 513, L127

\bibitem[{Sibthorpe {et~al.}(2018)Sibthorpe, Kennedy, Wyatt, Lestrade, Greaves,
  Matthews, \& Duchêne}]{sibthorpe+2018}
Sibthorpe, B., Kennedy, G.~M., Wyatt, M.~C., {et~al.} 2018, \mnras, 475, 3046

\bibitem[{Stapelfeldt {et~al.}(2004)Stapelfeldt, Holmes, Chen, Rieke, Su,
  Hines, Werner, Beichman, Jura, Padgett, Stansberry, Bendo, Cadien, Marengo,
  Thompson, Velusamy, Backus, Blaylock, Egami, Engelbracht, Frayer, Gordon,
  Keene, Latter, Megeath, Misselt, Morrison, Muzerolle, Noriega-Crespo,
  Van~Cleve, \& Young}]{stapelfeldt+2004}
Stapelfeldt, K.~R., Holmes, E.~K., Chen, C., {et~al.} 2004, \apjs, 154, 458

\bibitem[{Strubbe \& Chiang(2006)}]{strubbe+chiang2006}
Strubbe, L.~E. \& Chiang, E.~I. 2006, \apj, 648, 652

\bibitem[{Thébault \& Wu(2008)}]{thebault+wu2008}
Thébault, P. \& Wu, Y. 2008, \aap, 481, 713

\bibitem[{Thilliez \& Maddison(2016)}]{thilliez+maddison2016}
Thilliez, E. \& Maddison, S.~T. 2016, \mnras, 457, 1690

\bibitem[{Wyatt(2003)}]{wyatt2003}
Wyatt, M.~C. 2003, \apj, 598, 1321

\bibitem[{Wyatt(2005{\natexlab{a}})}]{wyatt2005}
Wyatt, M.~C. 2005{\natexlab{a}}, \aap, 433, 1007

\bibitem[{Wyatt(2005{\natexlab{b}})}]{wyatt2005b}
Wyatt, M.~C. 2005{\natexlab{b}}, \aap, 440, 937

\bibitem[{Wyatt {et~al.}(1999)Wyatt, Dermott, Telesco, Fisher, Grogan, Holmes,
  \& Piña}]{wyatt+1999}
Wyatt, M.~C., Dermott, S.~F., Telesco, C.~M., {et~al.} 1999, \apj, 527, 918

\end{thebibliography}
\end{document}